\documentclass[prl,preprint,preprintnumbers,superscriptaddress,showpacs]{revtex4}

\newcommand{\bs}   {\boldsymbol}
\newcommand{\mb}   {\mathbf}

\newcommand{\mcal} {\mathcal}
\newcommand{\imag} {\mathrm{i}}
\newcommand{\dd}   {\mathrm{d}}
\newcommand{\e}    {\mathrm{e}}
\newcommand{\bra}  {\langle}
\newcommand{\ket}  {\rangle}
\newcommand{\up}   {\uparrow}
\newcommand{\dn}   {\downarrow}
\newcommand{\s}    {\sigma}
\newcommand{\w}    {\omega}

\newcommand{\eps}  {\epsilon}

\usepackage{ulem}
\usepackage{amsmath}
\usepackage{amssymb}
\usepackage{latexsym} 
\usepackage{graphicx}

\begin{document}

\title{Electron-hole doping asymmetry of Fermi surface reconstructed in a simple Mott insulator}

\author{Yoshitaka Kawasugi} \email{kawasugi@riken.jp}
\affiliation{RIKEN, Wako, Saitama 351-0198, Japan}
\author{Kazuhiro Seki}
\affiliation{RIKEN, Wako, Saitama 351-0198, Japan}
\affiliation{RIKEN Advanced Institute for Computational Science (AICS), Kobe, Hyogo 650-0047, Japan}
\author{Yusuke Edagawa}
\affiliation{Department of Applied Physics, Waseda University, Tokyo 169-8555, Japan}
\author{Yoshiaki Sato}
\affiliation{RIKEN, Wako, Saitama 351-0198, Japan}
\author{Jiang Pu}
\affiliation{Department of Applied Physics, Waseda University, Tokyo 169-8555, Japan}
\author{Taishi Takenobu}
\affiliation{Department of Applied Physics, Waseda University, Tokyo 169-8555, Japan}
\author{Seiji Yunoki}
\affiliation{RIKEN, Wako, Saitama 351-0198, Japan}
\affiliation{RIKEN Advanced Institute for Computational Science (AICS), Kobe, Hyogo 650-0047, Japan}
\affiliation{RIKEN Center for Emergent Matter Science (CEMS), Wako, Saitama 351-0198, Japan}
\author{Hiroshi M. Yamamoto} \email{yhiroshi@ims.ac.jp}
\affiliation{RIKEN, Wako, Saitama 351-0198, Japan}
\affiliation{Research Center of Integrative Molecular Systems (CIMoS), Institute for Molecular Science, National Institutes of Natural Sciences, Okazaki, Aichi 444-8585, Japan}
\author{Reizo Kato}
\affiliation{RIKEN, Wako, Saitama 351-0198, Japan}
 
\begin{abstract}
It is widely recognised that the effect of doping into a Mott insulator is complicated and unpredictable, as can be seen by examining the Hall coefficient in high $T_{\rm c}$ cuprates.
The doping effect, including the electron-hole doping asymmetry, may be more straightforward in doped organic Mott insulators owing to their simple electronic structures.
Here we investigate the doping asymmetry of an organic Mott insulator by carrying out electric-double-layer transistor measurements and using cluster perturbation theory.
The calculations predict that strongly anisotropic suppression of the spectral weight results in the Fermi arc state under hole doping, while a relatively uniform spectral weight results in the emergence of a non-interacting-like Fermi surface in the electron-doped state. 
In accordance with the calculations, the experimentally observed Hall coefficients and resistivity anisotropy correspond to the pocket formed by the Fermi arcs under hole doping and to the non-interacting Fermi surface under electron doping.
\end{abstract}

\maketitle
\section*{Introduction}
Electron-hole doping asymmetry in Mott insulators has been one of the key questions related to the origin of superconductivity in the proximity of the insulating state \cite{Dagotto,Imada,Lee}.
To investigate this, one should prepare an exactly half-filled (zero-doped) Mott insulator and then inject or extract electrons, preferably by an electrostatic method, in the same sample.
However, it is difficult to perform such measurements on high-$T_{\rm c}$ cuprates for the following three reasons.
First, they are mostly either only hole-doped or electron-doped, and the crystallographic structures of the parent materials are often different. 
Second, they require a strong electric field to tune their band filling owing to their high half-filled carrier density (on the order of 10$^{15}$ cm$^{-2}$). 
Third, since they are charge-transfer-type insulators, electrons and holes are doped into different electronic orbitals, which may obscure the pure doping asymmetry.
In practice, the inverse of the Hall coefficient monotonically increases with increasing doping concentration with the sign changing across the zero-doping point \cite{Segawa}. 
This is a feature of band insulators rather than Mott insulators.

In contrast to the high-$T_{\rm c}$ cuprates, the organic Mott insulator $\kappa$-(BEDT-TTF)$_{2}$Cu[N(CN)$_{2}$]Cl \cite{Williams} (abbreviated to $\kappa$-Cl hereafter), where BEDT-TTF represents bis(ethylenedithio)tetrathiafulvalene, serves as an appropriate material for examining the electron-hole asymmetry of doped Mott insulators because of its relatively low carrier density ($\sim$1.8$\times$10$^{14}$ cm$^{-2}$) and its single electronic orbital nature. 
Indeed, electrons and holes are doped into the same $\pi$ electron band and thus $\kappa$-Cl can be modeled as a single band Hubbard model on an anisotropic triangular lattice (Fig.~\ref{fig.1}). 
Due to its high controllability of electronic states, $\kappa$-Cl has been studied intensively in terms of the bandwidth-controlled Mott transition between Mott insulating and superconducting states at half filling \cite{Ueno,Suda}.
The electrostatic method of carrier doping into organic Mott insulators with field-effect transistor (FET) structure has been implemented by the authors \cite{Kawasugi,Motoyama}. 
However, the limited doping concentration with the FET still forbids the observation of electron-hole asymmetry.

Here, by fabricating electric-double-layer transistors (EDLTs) based on thin single crystals of $\kappa$-Cl, we realised for the first time both electron and hole doping into the organic Mott insulator.
We measured the field effect on transport properties including the sheet resistivity, Hall coefficient and resistivity anisotropy to elucidate the doping anisotropy of the electronic state in the doped organic Mott insulator.
To provide further insights into the Fermi surface (FS), we calculated the single-particle spectral functions for an effective model of $\kappa$-Cl by using cluster perturbation theory (CPT) \cite{Gros1993,Senechal2000}.
The calculations predict that strongly anisotropic suppression of the spectral weight results in the Fermi arc state under hole doping, while a relatively uniform spectral weight results in the emergence of a non-interacting-like FS in the electron-doped state. 
In accordance with the calculations, the experimentally observed Hall coefficients and resistivity anisotropy correspond to the pocket formed by the Fermi arcs under hole doping and to the non-interacting FS under electron doping.

\section*{Results}
\subsection*{Resistivity}
We fabricated the EDLT devices by mounting an ion gel on a Hall-bar-shaped thin single crystal of $\kappa$-Cl and a Au side gate electrode (Figs. 2a and 2b).
First, we obtained the resistivity curve for the sheet resistivity $\rho$ at the temperature $T=220$ K (Fig.~\ref{fig.2}c).
As expected for a Mott insulator, a clear ambipolar field effect with a resistance peak at the gate voltage $V_{\rm g} \simeq -0.2$ V was observed.
The hysteresis and the leakage current remained small without any signature of a chemical reaction between the electrolyte and $\kappa$-Cl.
Note that the field effect was about 3\% at 220 K because thermally excited carriers in the bulk (about 100 molecular layers in a typical sample) dominated the electrical conduction owing to the low charge excitation gap (ca. 20 meV).
We also confirmed that the carrier tunability of the $\kappa$-Cl EDLT greatly exceeded that of SiO$_{2}$-based FETs by comparing these devices using the same $\kappa$-Cl crystal (Fig.~\ref{fig.s1}).
Provided that the mobilities of the FET and EDLT are equivalent, a gate voltage of +1 V in the EDLT corresponded to approximately 20\% electron doping (see Supplementary Note 1).
The resistivity curves at 220 K and the temperature dependences of the resistivity were reproducible after multiple temperature cycles (see Supplementary Note 2 and Fig.~\ref{fig.s2}).

With decreasing temperature, the ambipolar field effect became more distinct.
Figures ~\ref{fig.2}d and ~\ref{fig.2}e show the temperature dependence of the sheet resistivity under a negative gate voltage (hole doping) and positive gate voltage (electron doping), respectively.
Under hole doping, the activation energy significantly decreased to less than 1 meV at $V_{\rm g}$ = $-$1.35 V.
On the other hand, the doping of electrons tended to reduce the resistivity more effectively and metallic conductivity was observed at $V_{\rm g}>$1 V.
In a high-conductivity sample, we observed negative magnetoresistance accompanied with the upturn of resistivity at low temperature, which suggested a weak localisation effect (see Supplementary Note 3 and Fig.~\ref{fig.s3}).
Analysis of the weak localisation indicated that the mean free path and dephasing length considerably exceeded the distance between BEDT-TTF dimers indicating coherent transport.

\subsection*{Hall effect}
To observe a more significant difference between the electron- and hole-doped states, we measured the Hall coefficient $R_{\rm H}$, which indicates the FS topology.
The magnetic field $B$ dependence of the Hall resistance $R_{xy}$ at 30 K is shown in Fig.~\ref{fig.3}a ($R_{\rm H}$ is given by the slope).
Despite electron doping, the Hall coefficients were clearly positive and a simple estimation of the carrier density, $1/eR_{\rm H}$, gave about 0.8 holes per BEDT-TTF dimer.
Since the injected carriers were electrons ($\sim$0.2 electrons/dimer), the concentration of which was much lower than that of the observed holes, it suggests that dense hole carriers are delocalised by electron doping.
This is considered to be the doping-induced Mott transition previously observed in FET devices \cite{Kawasugi2}.

If we assume doping symmetry, a similar Hall effect is expected for the hole-doped side.
Namely, hole doping immediately collapses the Mott-Hubbard gap and $1/eR_{\rm H}$ corresponding to 1$-\delta$ ($\delta$: electron doping concentration, which is negative for hole doping) holes per BEDT-TTF dimer should be observed.
However, as shown in Fig.~\ref{fig.3}a, the Hall coefficient on the hole-doped side was about three times larger than that on the other side, and $1/eR_{\rm H}$ gave about 0.3 holes per BEDT-TTF dimer.
Although $1/eR_{\rm H}$ increased with increasing absolute value of the gate voltage, the rate of increase was low (only 7\% difference between $V_{\rm g}$ = $-$0.8 V and -1.2 V).
Surprisingly, under hole doping, $1/eR_{\rm H}$ per dimer appears to be neither 1$-\delta$ (delocalised hole carriers + injected hole carriers) nor $-\delta$ (only injected hole carriers), indicating that the localised carriers were partially delocalised by excess hole doping as shown in Fig.~\ref{fig.3}c.
These results indicate that the FS topology is significantly different between the electron- and hole-doped states.
For the gate voltage dependence of the Hall mobility, see Supplementary Note 4 and Fig.~\ref{fig.s5}.
The temperature dependence of $R_{\rm H}$ (Fig.~\ref{fig.3}b) shows that the $R_{\rm H}$ asymmetry was maintained up to about 50 K, above which the Hall effect at the doped surface was obscured by the thermally excited carriers in the bulk.
Note that increase in $R_{\rm H}$ at 30 K upon cooling is due to the decrease of the thermally excited carriers in the bulk (see Supplementary Note 5 and Fig.~\ref{fig.s6}). 
%
%
The data below 25 K contain ambiguity owing to the non-ohmic behaviour (see Supplementary Note 6 and Fig.~\ref{fig.s4}).

What is the origin of $R_{\rm H}$ under hole doping?
As shown in Fig.~\ref{fig.1}, $\kappa$-Cl forms essentially an anisotropic triangular lattice resulting in an ellipsoidal non-interacting FS which is folded by the zone boundary.
In a Fermi liquid with a single type of carrier, $1/eR_{\rm H}$ denotes the carrier density corresponding to the volume enclosed by the FS (Luttinger's theorem \cite{Luttinger}).
Under electron doping, $1/eR_{\rm H}$ is close to the carrier density corresponding to the non-interacting ellipsoidal FS (known as the $\beta$-orbit in the study of Shubnikov-de Haas oscillation \cite{Mielke}).
By contrast, $1/eR_{\rm H}$ under hole doping somehow appears to correspond to the volume bounded by the lenslike closed portion of the FS (known as the $\alpha$-orbit).
The doping dependences of the carrier density enclosed by the $\alpha$- and $\beta$-orbits assigned for a $\kappa$-BEDT-TTF salt \cite{Oshima} are plotted as reference in Fig.~\ref{fig.3}c.

\subsection*{Single-particle spectral functions}
To verify this $R_{\rm H}$ asymmetry, we calculated the single-particle spectral functions of the Hubbard model on an anisotropic triangular lattice ($t'/t$ = $-$0.8, $U/t$ = 7) at 30 K using CPT and the results are shown in Fig.~\ref{fig.4}.
First, the Mott insulating state was reproduced at half filling, where the energy gap opened at all the k-points (Figs. 4b and 4e).
When electrons were doped ($\delta$ = +0.17), the FS appeared as shown in Figs. 4c and 4f.
The FS topology was the same as that of the non-interacting case, although the shape and the spectral weight were not exactly the same.
In this state, the value of $R_{\rm H}$ should be close to that of the Fermi liquid state, as suggested by the experiments.
On the other hand, the FS under hole doping ($\delta$ = $-$0.17) was very different from the non-interacting FS (Figs. 4a and 4d).
The spectral weight near the Z point was strongly suppressed.
Namely, arcs around the X point and pseudogaps near the Z point appeared.
The partial disappearance of FS by the pseudogap is similar to the Fermi arc state observed in lightly hole-doped cuprates \cite{Marshall, Norman, Yoshida}.
In this state, Luttinger's theorem is seemingly violated and the value of $R_{\rm H}$ is no longer simply estimated.
However, considering the first Brillouin zone (BZ) of $\kappa$-Cl, one can see that the Fermi arcs are folded and form closed lenslike orbits.
These orbits originate from the $\alpha$-orbit mentioned above and correspond to the observed $R_{\rm H}$ in the hole-doped states.
These results imply that $R_{\rm H}$ is predominantly governed by quasi-particles with relatively long lifetime (bright points of the spectral function in the reciprocal space in Fig.~\ref{fig.4}).
Therefore, the asymmetry of the FS topology was clearly observed via $R_{\rm H}$.

\subsection*{Anisotropy of the resistivity}
The partial disappearance of the FS can also be confirmed by the anisotropy in the carrier conduction. 
The strongly suppressed parts of the FS under hole doping should affect strongly the carrier conduction along the $c$-axis.
Indeed, our CPT calculations predict that the resistivity is more anisotropic in the hole-doped state than in the electron-doped state (see Supplementary Note 7 and Fig.~\ref{fig.opt}).
We measured the resistivity anisotropy as shown in Fig.~\ref{fig.5}.
The crystallographic axes were determined by the crystal shape and the sign of the Seebeck coefficient at room temperature.
At 220 K where the thermally excited carriers in the non-doped bulk dominates the conductivity, the $c$-axis resistivity $\rho _{c}$ was slightly higher than the $a$-axis resistivity $\rho_a$ in both electron and hole doping.
With decreasing temperature, $\rho _{c} /\rho _{a}$ increased under hole doping while it decreased and remained near 1 in the non-doped and electron-doped states as shown in Fig.~\ref{fig.5}b.
The contour plot of $\rho_{c} / \rho_{a}$ under various gate voltage and temperature (Fig.~\ref{fig.5}c) clearly exhibited that the resistivity in the hole-doped state was particularly anisotropic, in good qualitative agreement with our CPT calculations.

\section*{Discussion}
The doping asymmetry is expected from the particle-hole asymmetric non-interacting band structure, as shown in Fig.~\ref{fig.1}c.
The energy bands along the Z-M axis are very flat and the van Hove singularity (vHs) lies below the FS at half filling.
With the doping of holes, the FS approaches the vHs and the effect of the interaction is expected to be enhanced.
Indeed, substantial suppression of the spectral weight of the FS along the Z-M axis, on which the van Hove critical points lie, is observed (Fig.~\ref{fig.4}a).
By contrast, the FS departs from the vHs with the doping of electrons, resulting in a weaker interaction effect and a more non-interacting-like FS.
It is also indicated that the spin fluctuation is stronger in the hole-doped state because of the vHs.
If a superconducting state could be induced by further hole doping, the transition temperature and pairing symmetry are of great interest since the hole-doped state is substantially different from the metallic state created by controlling the bandwidth at half filling.

To summarise, without gate voltage, $\kappa$-Cl is a Mott insulator due to electron interaction and high commensurability between the hole density and periodic potential (1 hole / 1 site).
When holes or electrons are doped, the commensurability is reduced and the effect of electron interaction is weakened, resulting in carrier delocalisation.
However, the effect of interaction remains strong at specific k-points where the energy dispersion is relatively flat.
Owing to the particle-hole asymmetric band structure, the pseudogaps appeared only near the van Hove critical points under hole doping.
In contrast to the calculations, our measurements did not observe the true metallic state (${\rm d}\rho /{\rm d}T > 0$) at low temperature.
This is probably due to the effect of localisation caused by finite random potential at the surface \cite{Anderson1958,Anderson1979}, as the negative magnetoresistance indicated.
Improvement of the surface roughness and cleanness may realise metallic phases under both dopings.

Thus, the doping asymmetry in a Mott insulator was demonstrated by a simple process in an organic Mott EDLT.
The above results show that organic Mott insulators are good model materials for Mott physics even in the case of band-filling control, and the electronic state in a simple Mott insulator may be predictable in the framework of CPT.
Since molecular conductors consist of various molecules with different arrangements, the effect of doping into materials with different FS topologies is of interest in future experiments.
Furthermore, the presence/absence of the superconducting phase under doping and its doping asymmetry are also intriguing.

\section*{Methods}
\textbf{Sample preparation.}
The source, drain, and gate electrodes (18 nm-thick-Au) were patterned on a polyethylene naphthalate substrate (Teonex Q65FA, Teijin DuPont Films Japan Limited) using photolithography.
A thin ($\sim$100 nm) single crystal of $\kappa$-(BEDT-TTF)$_{2}$Cu[N(CN)$_{2}$]Cl (abbreviated to $\kappa$-Cl hereafter) was electrochemically synthesized by oxidizing BEDT-TTF (20 mg) dissolved in 50 ml of 1,1,2-trichloroethane (10\% v/v ethanol) in the presence of TPP[N(CN)$_{2}$] (TPP = tetraphenylphosphonium, 190 mg), CuCl (60 mg) and TPP-Cl (100 mg).
After applying current of 8 $\mu$A for 20 h, a thin crystal was transferred into 2-propanol with a pipette and was guided on top of the PEN substrate.
After the substrate was removed from the 2-propanol and dried, the $\kappa$-Cl crystal was shaped into a Hall-bar using a pulsed laser beam with a wavelength of 532 nm.
Typical dimensions of the Hall bar sample were approximately 15 $\mu$m (width) $\times$ 40 $\mu$m (length) $\times$ 100 nm (thickness).
The EDLT devices were completed by mounting an ion gel (or ionic liquid) on the Hall-bar-shaped thin single crystal of $\kappa$-Cl and a Au side gate electrode.
The basic techniques, such as the electrochemical synthesis of thin single crystals and the laser shaping, are common to our previous FET devices \cite{Kawasugi2}.
As gate electrolytes, we employed poly(vinylidene fluoride-co-hexafluoropropylene) [PVDF-HFP] with 58\% w/w 1-butyl-3-methylimidazolium tetrafluoroborate [BMIM-BF$_{4}$] (samples \#1$\sim$4), N,N-diethyl-N-methyl-N-(2-methoxyethyl) ammonium bis(trifluoromethanesulfonyl)imide [DEME-TFSI] (sample \#5 in Fig.~\ref{fig.s1}), polymethyl methacrylate [PMMA] with 57\% w/w DEME-TFSI (sample \#6 in Fig.~\ref{fig.s3}).
The PVDF-HFP ion gel \cite{Lee2} was spin-coated from an acetone solution in which the weight ratio between the polymer, ionic liquid, and solvent was 1:1.3:7.7.
Although the ionic conductivity was the greatest in the liquid state, the large thermal stress usually broke the $\kappa$-Cl crystal at a low temperature ($T <$150 K).
Thinning of the gate electrolytes by gelation reduced the thermal stress at low temperatures, allowing the successful cooling of the samples down to 2 K.
The area of the gate electrode was more than 100 times larger than that of the $\kappa$-Cl channel in order to apply the gate voltage effectively.

\textbf{Transport measurements.}
The transport measurements were performed using a Physical Property Measurement System (Quantum Design).
The temperature and magnetic field normal to the substrate were controlled in the range of 2-300 K (cooling rate: 2 K/min) and $\pm$8 Tesla, respectively.
When the gate voltage was varied, the samples were warmed to 220 K.
In the Hall measurements, the magnetic field was swept in the range of $\pm$8 Tesla at a constant temperature and the forward and backward data were averaged to eliminate the small linear voltage drift.

Although it is difficult to experimentally determine the distribution of field-induced charge along the out-of-plane direction, we consider that the field-induced carriers predominantly exist in a single conducting molecular sheet \cite{Kawasugi2}.
Recently, the quantum oscillations in the doped surface of the Dirac fermion system $\alpha$-(BEDT-TTF)$_{2}$I$_{3}$ \cite{Tajima} indicated that only two conducting layers from the surface were doped, in which the carrier density in the second layer was only 1/7 of that in the first layer.
The above assumption is also likely to apply to $\kappa$-type BEDT-TTF salts, where a shorter screening length is expected.

\textbf{Hubbard model on an anisotropic triangular lattice.}
To examine the electron correlation effects in the organic doped Mott insulator, 
we study the single-band Hubbard model defined on an anisotropic triangular lattice,  
which is known as the simplest model for $\kappa$-type BEDT-TTF salts~\cite{Kino1996,Fukuyama2006,Powell2006,Nakamura2009}. 
The model is defined by the Hamiltonian, 
\begin{equation}
  \label{eq.Ham}
  \hat{H} = - \sum_{\bra ij \ket ,\s} t_{ij} \left( \hat{c}_{i\s}^\dag \hat{c}_{j \s} + {\rm H. c.} \right) 
  + U \sum_{i} \hat{n}_{i \up} \hat{n}_{i \dn} 
  - \mu \sum_{i \s} \hat{n}_{i \s},
\end{equation}
where $\hat{c}_{i\s}^\dag$ creates an electron on site $i$ with spin $\s(=\up,\dn)$ 
and $\hat{n}_{i \s} = \hat{c}^\dag_{i \s} \hat{c}_{i \s}$. 
$t_{ij}$ is the transfer integral between the neighboring sites $i$ and $j$, 
$U$ is the on-site Coulomb repulsion, and 
$\mu$ is the chemical potential. 
Here, 
a single site consists of a single dimer and 
a unit cell contains two dimers with different orientation (Fig.~\ref{fig.lattice}a).
The transfer integral between the different (same) dimers is given as $t_{ij} = t$ ($t'$). 

The series of $\kappa$-type BEDT-TTF salts serves as an ideal 
realization of the Hubbard model of Eq.~(\ref{eq.Ham}). 
The anisotropy $|t'/t|$ can be changed depending on substituted anions~\cite{Shimizu2003,Kandpal2009,Kanoda2006,Nakamura2009}.  
The Coulomb repulsion $U$ is estimated to be comparable to~\cite{Kandpal2009, Komatsu1996} 
or larger than~\cite{Nakamura2009} the band width, 
implying that the strongly correlated electronic state is realized at ambient pressure. 
Applied pressure~\cite{Kanoda2006,Kurosaki2005,Oike2015} or bending strain on thin films~\cite{Suda2014} 
can control the electron correlation parameter $t/U$.
Moreover, the chemical potential $\mu$ can be 
tuned efficiently by applied gate voltage in 
the EDLT, allowing for investigation of the electronic structure 
of electron- and hole-doped Mott insulators on equal footing. 
The series of organic charge transfer salts is indeed 
the highly controllable counterpart of strongly correlated  
transition-metal compounds~\cite{ZSA}. 
In the main text, we have focused on a typical parameter set for $\kappa$-type BEDT-TTF salts of 
$t'/t = -0.8$, $U/t = 7$, and $t = 55$ meV. 
We have also considered the parameter set relevant for $\kappa$-Cl of 
$t'/t = -0.44$, $U/t = 5.5$, and $t = 65$ meV~\cite{Kandpal2009}, and 
found that the Fermi arc appears for hole-doped case also in this parameter set (Fig.~\ref{fig.4another}).    
The non-interacting tight-binding band structure, the density of states, 
and the FS at half filling for this parameter set are shown in Figs.~\ref{fig.lattice}c and ~\ref{fig.lattice}d.

\begin{figure}[htbp]
  \begin{center}
    \includegraphics{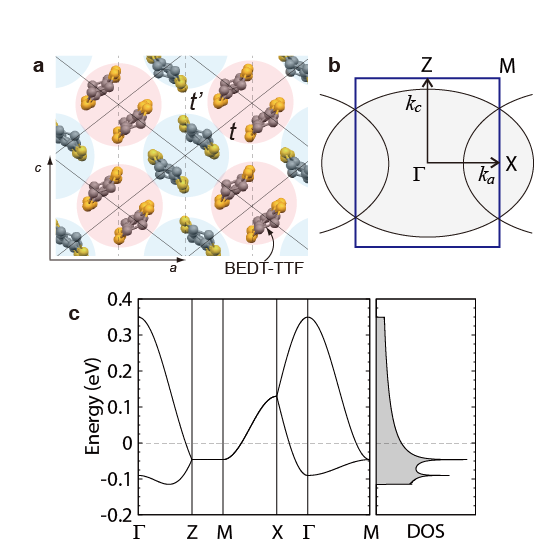}
  \caption{\textbf{Molecular arrangement and electronic structure of $\kappa$-Cl without electron correlation. a,} Molecular arrangement of the BEDT-TTF layer and the effective anisotropic triangular lattice. $t$ is the nearest-neighbor hopping (solid line) and $t'$ is the next-nearest-neighbor hopping (dashed line) between the BEDT-TTF dimers. Two differently oriented dimers are indicated by red and blue shades. \textbf{b,} Schematic of Fermi surface without electron interaction. The 1st BZ of $\kappa$-Cl is indicated by a solid square. \textbf{c,} Tight-binding band structure and density of states (DOS) for $t'/t$ = $-$0.8 with $t$ = 55 meV. Here, $\Gamma = (0,0)$, Z$ = (0,\pi/c)$, M$ = (\pi/a,\pi/c)$, and X$ = (\pi/a,0)$. The Fermi energy is denoted by dashed lines.}
  \label{fig.1}
  \end{center}
\end{figure}

\begin{figure}[htbp]
  \begin{center}
    \includegraphics{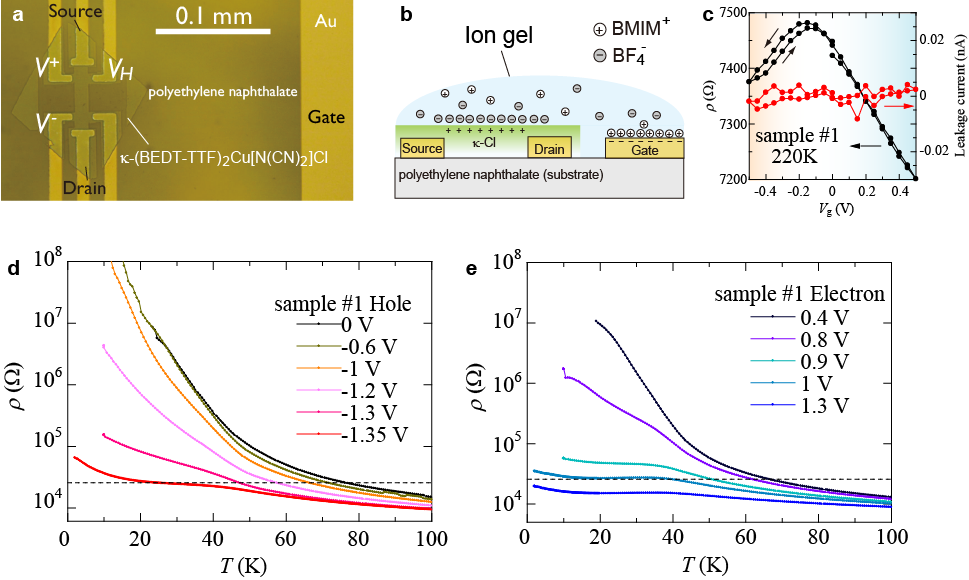}
  \caption{\textbf{EDLT device and field effect on sheet resistivity. a,b,} Optical top view \textbf{(a)} and schematic side view \textbf{(b)} of the EDLT device. The $\kappa$-Cl crystal is shaped into a Hall-bar using a pulsed laser beam. \textbf{c,} Gate voltage dependence of the sheet resistivity at 220 K. \textbf{d,e,} Temperature dependence of the sheet resistivity under hole doping \textbf{(d)} and electron doping \textbf{(e)}. The dashed lines denote $h/e^{2}\sim$ 25.8 k$\Omega$. The gate voltage was consecutively applied in the order of appearance, namely: 0, $-$0.6, $-$1, $-$1.2, $-$1.3 and $-$1.35 V for hole doping and 0.4, 0.8, 0.9, 1 and 1.3 V for electron doping.}
    \label{fig.2}
  \end{center}
\end{figure}

The strongly correlated Hubbard model at half filling or the Heisenberg model 
on the anisotropic triangular lattice have been 
investigated theoretically and the ground state phase diagrams 
with various phases including metal, antiferromagnetic insulator, 
spiral, and spin liquid ground states 
have been obtained~\cite{Yunoki2006,Kyung2006,Sahebsara2006,Watanabe2008,Tocchio2009,Tocchio2013,Yamada2014,Laubach2015}. 
Most of these previous calculations on the Hubbard model 
at half filling have been carried out with $t'/t >0$. 
However, the sign of $t'$ is crucial 
when it comes to the electron-hole asymmetry of carrier doping 
because the sign determines whether the vHs 
appears in hole-doped side or electron-doped side. 
The correct sign is $t'/t<0$~\cite{Nakamura2009}  
and the vHs appears in the hole-doped side (see Fig.~\ref{fig.1}c and Fig.~\ref{fig.lattice}c).

\textbf{Unit cell and Brillouin zone.} 
In Figs.~\ref{fig.4} and ~\ref{fig.4another}, we show the single-particle spectral functions and the FS 
in the BZ of the unit cell consisting of two sites 
in the real space (see blue shaded region in Fig.~\ref{fig.BZ}a). 
The BZ is relevant for the realistic structure of 
the $\kappa$-type BEDT-TTF salts and therefore the results are 
directly compared with those obtained by first-principles 
band-structure calculations~\cite{Kandpal2009,Nakamura2009}.  
However, the primitive unit cell of the single-band 
Hubbard model in Eq.~(\ref{eq.Ham}) on the anisotropic triangular lattice 
consists of one site (see red shaded region in Fig.~\ref{fig.BZ}a),  
unless a spontaneous symmetry breaking such as 
the N\'{e}el order occurs. 
Therefore the theoretical calculations for the Hubbard model on the anisotropic triangular lattice  
are usually performed with 
this primitive unit cell~\cite{Kyung2006,Sahebsara2006,Watanabe2008,Tocchio2009,Tocchio2013,Yamada2014,Laubach2015,Kang2011}.  
In order to avoid possible confusions caused by 
the different choice of the unit cell, we describe 
the relation between the two BZs of the one-site unit cell and the two-site unit cell. 

Figure~\ref{fig.BZ}a shows the anisotropic triangular lattice 
with regular-triangular plaquettes. 
The primitive translational vectors for the one (two)-site unit cell are 
given by $\{{\mb{e}}_1, {\mb{e}}_2\}$ ($\{{\mb{a}}, {\mb{c}}\}$), 
and the corresponding BZ is diamond (rectangular) shaped, as shown in Fig.~\ref{fig.BZ}c.   

We also show in Fig.~\ref{fig.BZ}b 
the lattice which is topologically equivalent to the 
triangular lattice in Fig.~\ref{fig.BZ}a 
but forms the square plaquettes with diagonal lines. 
Notice that this square topology of the lattice is very often used 
for study of the Hubbard model on the anisotropic triangular lattice~\cite{Kyung2006,Sahebsara2006,Watanabe2008,Tocchio2009,Tocchio2013,Kang2011}. 
The primitive translational vectors for the one (two)-site unit cell 
are again given by $\{{\mb{e}}_1, {\mb{e}}_2\}$ ($\{{\mb{a}}, {\mb{c}}\}$).  
The BZ of the one (two)-site unit cell is now square (diamond) shaped, as shown in Fig.~\ref{fig.BZ}d.   

Let us now consider the basis-transformation formula between the bases 
$\{{\mb{a}}, {\mb{c}}\}$ and $\{{\mb{e}}_1, {\mb{e}}_2\}$, where  
the former is a set of the primitive translational vectors of 
the realistic structure of the $\kappa$-type BEDT-TTF salts, 
while the latter is that of the single-band Hubbard model. 
It is easily found from Fig.~\ref{fig.BZ}b that 
they are related by the $\pm \pi/4$ rotation and 
the scale transformation. 
To be concrete, we consider a vector 
${\mb{r}} = r_a {\mb{a}} + r_c {\mb{c}} = r_1 {\mb{e}}_1 + r_2 {\mb{e}}_2$ 
in the real space. It is easily shown that 
the coordinate $(r_a, r_c)$ is related to the coordinate $(r_1, r_2)$ as 
\begin{eqnarray}
  \left(
  \begin{array}{c}
    r_a \\ 
    r_c 
  \end{array}
  \right)
  = 
  \frac{1}{\sqrt{2}}
  \left(
  \begin{array}{cc}
    \cos{\frac{\pi}{4}} & -\sin{\frac{\pi}{4}} \\
    \sin{\frac{\pi}{4}} &  \cos{\frac{\pi}{4}} 
  \end{array}
  \right) 
  \left(
  \begin{array}{c} 
    r_1 \\
    r_2    
  \end{array}
  \right)
  =
  \frac{1}{2}
  \left(
  \begin{array}{c} 
    r_1 - r_2\\
    r_1 + r_2    
  \end{array}
  \right).
\end{eqnarray}

Similarly, for a wave vector 
${\mb{k}} 
= k_a {\mb{g}}_a + k_c {\mb{g}}_c
= k_1 {\mb{g}}_1 + k_2 {\mb{g}}_2$, 
where $\{2\pi{\mb{g}}_a, 2\pi{\mb{g}}_c \}$ ($\{2\pi{\mb{g}}_1, 2\pi{\mb{g}}_2 \}$) 
is the set of the reciprocal vectors of 
$\{{\mb{a}}, {\mb{c}} \}$ ($\{{\mb{e}}_1, {\mb{e}}_2 \}$),   
($k_{a}, k_{c})$ is related to ($k_{1}, k_{2})$ as
\begin{eqnarray}
  \left(
  \begin{array}{c}
    k_a \\ 
    k_c 
  \end{array}
  \right)
  = 
  \sqrt{2}
  \left(
  \begin{array}{cc}
    \cos{\frac{\pi}{4}} & -\sin{\frac{\pi}{4}} \\
    \sin{\frac{\pi}{4}} &  \cos{\frac{\pi}{4}} 
  \end{array}
  \right)
  \left(
  \begin{array}{c} 
    k_1 \\
    k_2    
  \end{array}
  \right) 
  =
  \left(
  \begin{array}{c} 
    k_1 - k_2\\
    k_1 + k_2   
  \end{array}
  \right).
  \label{def.k1k2}
\end{eqnarray} 
For example, 
$(k_a,k_c) = (0,0), (0, \pi), (\pi,\pi) $, and $(\pi,0)$ in the $\{{\mb{g}}_a,{\mb{g}}_c\}$ basis 
(i.e., $\Gamma$, Z, M, and X points in Fig.~\ref{fig.BZ}c) correspond to 
$(k_1,k_2) = (0,0), (\pi/2, \pi/2), (\pi,0) $, and $(\pi/2, -\pi/2)$ in the $\{{\mb{g}}_1,{\mb{g}}_2\}$ basis (Fig.~\ref{fig.BZ}d), 
respectively. 
Finally, we note that the rectangular BZ in Fig.~\ref{fig.BZ}c corresponds 
to the antiferromagnetic BZ of the square lattice in Fig.~\ref{fig.BZ}d (blue shaded regions).

\textbf{Cluster perturbation theory.}
To examine the electron-hole asymmetry of the FS, 
the calculation of single-particle Green's functions is required. 
For this purpose, we employ the CPT~\cite{Senechal2000}.  
The CPT is a theory of strong coupling expansion for the single-particle Green's function and 
allows us to examine the single-particle excitations including 
quasi-particle excitations, Mott gap, and even pseudogap phenomena 
in strongly correlated electron systems~\cite{Senechal2004,Kang2011,Kohno2012}. 

As shown in the following, the CPT requires 
the fully interacting single-particle Green's function (or the self-energy~\cite{Potthoff2003,Dahnken2004}) 
of open-boundary clusters to obtain the single-particle Green's function of the lattice.  
The basic notion of the scheme was introduced in the earlier work by Gros and Valent{\'i}~\cite{Gros1993}. 

Let us outline the method of CPT. 
First, we divide the whole lattice on which the model Hamiltonian $\hat{H}$ is defined into identical 
finite-size clusters, each of which consists of $L_{\rm c}$ sites. 
We shall denote 
the Hamiltonian of a cluster as $\hat{H}_{\rm c}$ and 
the inter-cluster hopping as $\hat{\mcal{T}}$. 
Here, $\hat{H}_{\rm c}$ is given in the right-hand side of Eq.~(\ref{eq.Ham}) 
but it is defined on a $L_{\rm c}$-site cluster with open boundary conditions. 
Next, the single-particle Green's function $\bs{G}_{\rm c}$ of 
the cluster Hamiltonian $\hat{H}_{\rm c}$ is calculated 
by the exact diagonalization method. 
Applying the strong coupling expansion 
with respect to the inter-cluster hopping $\hat{\mcal{T}}$, 
the single-particle Green's function 
is given as $\tilde{\bs{G}} = \left(\bs{G}_{\rm c}^{-1}-\bs{\mcal{T}} \right)^{-1}$, 
where $\bs{\mcal{T}}$ is the matrix representation of $\hat{\mcal{T}}$~\cite{Senechal2002}. 
By restoring the translational invariance, 
which is broken in $\tilde{\bs{G}}$ due to the partitioning of the lattice, 
we obtain the single-particle Green's function $G$ of the whole lattice  
for momentum $\mb{k}$ and complex frequency $z$ as 
\begin{equation} 
  G(\mb{k},z) 
  = \frac{1}{L_{\rm c}}
  \sum_{i, j} 
  \left[
    \left(
      {\bs{G}_{\rm c}^{-1}(z)} - \bs{\mcal{T}}(\mb{k})
    \right)^{-1}
  \right]_{ij} 
  \e^{-\imag \mb{k} \cdot (\mb{r}_i - \mb{r}_j)},
  \label{eq.cpt}
\end{equation}
where $\mb{r}_i$ is the position of the $i$-th site within the cluster~\cite{Senechal2002}.  

The cluster Green's function matrix $\bs{G}_{\rm c}$ is given as  
\begin{equation}
  \label{lehmann}
  \bs{G}_{{\rm c}} (z) =  \bs{G}^+_{{\rm c}}(z) + \bs{G}^-_{{\rm c}}(z) 
\end{equation}
with 
\begin{eqnarray}
  \label{eq.Ge}
  {G}^+_{{\rm c},ij}(z) &=& \sum_{s=0}^{s_{\rm max}} \frac{\e^{-\beta E_s}}{Z} \bra \Psi_s | \hat{c}_{i\s}  \left(z - \hat{H}_{\rm c} + E_s \right)^{-1}  \hat{c}_{j\s}^\dag | \Psi_s \ket 
\end{eqnarray}
and 
\begin{eqnarray}
  \label{eq.Gh}
  {G}^-_{{\rm c},ij}(z) &=& \sum_{s=0}^{s_{\rm max}} \frac{\e^{-\beta E_s}}{Z} \bra \Psi_s | \hat{c}_{j\s}^\dag \left(z + \hat{H}_{\rm c} - E_s  \right)^{-1} \hat{c}_{i\s} | \Psi_s \ket, 
\end{eqnarray}
where $\beta = 1/k_{\rm B}T$ with $T$ being the temperature, 
$|\Psi_s \ket $ is the $s$-th eigenstate of $\hat{H}_{\rm c}$ with its eigenvalue $E_s$ 
in ascending order, and 
$Z = \sum_{s=0}^{s_{\rm max}} \e^{-\beta E_s}$ is the partition function of the cluster. 
The sum over eigenstates is truncated by summing up to the $s_{\rm max}$-th eigenstate. 
The upper bound $s_{\rm max}$ of the sum is determined 
so as to satisfy that $\e^{-\beta E_{s_{\rm max}}}/\e^{-\beta E_0} < \eps$, 
where $E_0$ is the ground-state energy and $\eps$ is a small real number. 
For the finite temperature calculations at $T = 100$~K,  
we set $\eps = 10^{-5}$, in which for example 
the lowest $s_{\rm max} \sim 1400$ eigenstates among $4^{12}$ eigenstates in total are 
included for the hole-doped case with $L_{\rm c} = 12$. 

The inter-cluster hopping between different clusters in Eq.~(\ref{eq.cpt}) is represented as
\begin{equation}
  \label{eq.Tij}
  \mcal{T}_{ij}(\mb{k}) = -t_{ij} \sum_{m_1,m_2} 
  \exp{
    \left[ \imag \mb{k} \cdot 
      \left(m_1 \mb{X}_1 + m_2 \mb{X}_2 \right)
    \right]
  }, 
\end{equation}
where 
$t_{ij}$ is given in Eq.~(\ref{eq.Ham}), 
$\mb{X}_1$ and $\mb{X}_2$ are the translational vectors of the clusters, and 
$m_1\mb{X}_1 + m_2 \mb{X}_2$ ($m_1$ and $m_2$: integer) indicates a relative position 
of two different clusters, site $i$ being located in one cluster and site $j$ in the other cluster. 
For the 12-site cluster, the translational vectors are  
$\mb{X}_1 = 2.5\mb{a}+0.5\mb{c}$ and $\mb{X}_2 = 0.5\mb{a} + 2.5\mb{c}$, and 
$t_{ij}$ in Eq.~(\ref{eq.Tij}) is finite only when 
$(m_1,m_2) = (\pm1,0)$, $(0,\pm1)$, and $(\pm1,\mp1)$, indicating 
the inter-cluster hopping between a cluster and its neighboring six clusters 
(see Fig.~\ref{fig.lattice}a). 
For the 16-site cluster, the translational vectors are 
$\mb{X}_1 = 2\mb{a}+2\mb{c}$ and $\mb{X}_2 = -2\mb{a} + 2\mb{c}$, and 
$t_{ij}$ in Eq.~(\ref{eq.Tij}) is finite only when 
$(m_1,m_2) = (\pm1,0)$, $(0,\pm1)$, and $(\pm1,\pm1)$ (see Fig.~\ref{fig.lattice}b).  

Note that in principle a cluster with a larger size gives better results in the CPT 
because the electron correlation effects including spatial fluctuations 
within a cluster are fully taken into account. 
We have shown in the main text and Fig.~\ref{fig.4another} the single-particle spectral functions using 
the 12-site cluster at finite temperatures. 
We also show in Figs.~\ref{fig.Akw1} and ~\ref{fig.Akw2} the single-particle spectral functions for the 16-site cluster at zero temperature (see also Supplementary Note 8), 
where $s_{\rm max} = 0$ in Eqs.~(\ref{eq.Ge}) and (\ref{eq.Gh}). 
The results clearly demonstrate that 
the electron-hole asymmetric reconstruction of the FS 
is observed irrespectively of the cluster sizes used. 
As a technical point, 
we have made use of the $C_{2v}$ symmetry of the 12-site and 16-site clusters 
to reduce computational costs for solving the eigenvalue problem of $\hat{H}_{\rm c}$ and  
for calculating the cluster single-particle Green's functions.

\textbf{Single-particle spectral function.}
As shown in Fig.~\ref{fig.lattice}a, the partitioning of the triangular lattice into the 12-site clusters 
does not respect the $C_{2v}$ symmetry (for lack of the $\sigma_v$ reflection symmetry) 
and the two-site unit cell structure (as $\mb{X}_1$ and $\mb{X}_2$ connect differently oriented dimers) 
of the dimer model on the anisotropic triangular lattice. 
However, these can be restored by 
averaging the single-particle Green's functions over four different momenta, i.e., 
\begin{equation}
  \bar{G}(\mb{k},\w) = \frac{1}{4} 
  \left[ 
    G(k_a,k_c,\w) + G(-k_a,k_c,\w) + G(k_a,k_c+\frac{2\pi}{c},\w) + G(-k_a,k_c+\frac{2\pi}{c},\w) 
  \right].
  \label{eq.Gbar}
\end{equation}
This is because the average of the Green's functions between $\mb{k} = (k_a,k_c)$ and $(-k_a,k_c)$ 
restores the $\sigma_{v}$ reflection symmetry and the addition of the Green's functions with wave 
vectors shifted by a primitive reciprocal vector, i.e., $\mb{k} = (\pm k_a,k_c+2\pi/c)$, 
recovers the two-site unit cell structure in the BZ. 
Therefore, $\bar{G}(\mb{k},\w)$ given in 
Eq.~(\ref{eq.Gbar}) provides 
the single-particle Green's function relevant for the Hubbard model 
defined on the anisotropic triangular lattice even with the 12-site cluster. 

On the other hand, partitioning of the lattice into the 16-site clusters respects 
both the $C_{2v}$ symmetry and the two-site unit cell structure (see Fig.~\ref{fig.lattice}b). 
Therefore, the single-particle Green's function with 16-site cluster 
is calculated in a standard way of the CPT for multi-band models described in~\cite{Senechal2002}. 

Using $\bar{G}(\mb{k},\w)$, the single-particle spectral function is calculated as  
\begin{equation}
  A(\mb{k},\w) = -\frac{1}{\pi} {\rm Im} \bar{G}(\mb{k},\w + \imag \eta),  
\end{equation} 
where a small imaginary part $\imag\eta$ of the complex frequency 
gives the Lorentzian broadening of the spectra. 
The FS is determined with the single-particle spectral function at zero energy, 
$A(\mb{k},0)$. 
We set the Lorentzian broadening factor 
$\eta/t = 0.2$ for the single-particle spectral functions and  
$\eta/t = 0.15$ for the FS calculation.  
The single-particle spectral functions and FS are shown in Figs. 4 , 16 and 17 ($t'/t$ = $-$0.8, $U/t$ = 7 and $t$ = 55 meV), and in Fig.~\ref{fig.4another} ($t'/t$ = $-$0.44, $U/t$ = 5.5 and $t$ = 65 meV).

\section*{Data Availability Statements}
The data that support the findings of this study are available from the corresponding authors upon request.
 
\section*{Acknowledgments}
We would like to acknowledge Teijin DuPont Films Japan Limited for providing the PEN films.
Computations have been done using HOKUSAI facility of Advanced Center for Computing and Communication at RIKEN.
This work was supported by MEXT and JSPS KAKENHI (Grant Nos. JP16H06346, 15K17714, 26102012 and 25000003), JST ERATO, and MEXT Nanotechnology Platform Program (Molecule and Material Synthesis).

\section*{Author contributions}
Y.K. and K.S. contributed equally to this work. 
Y.K. performed the planning, sample fabrication, cryogenic transport measurements and data analyses.
K.S. and S.Y. performed all CPT calculations and data analyses.
Y.E., J.P. and T.T. developed the ion gel for $\kappa$-Cl EDLT, performed transfer curve analyses and improved the device performance.
Y.S. grew $\kappa$-Cl single crystals of high quality and performed magnetoresistance data analyses.
Y.K., K.S. and H.M.Y. wrote the manuscript.
T.T., S.Y., H.M.Y. and R.K. supervised the investigation.
All authors commented on the manuscript.

\section*{Additional information}
Supplementary information is available in the online version of the paper. 
Reprints and permissions information is available online at www.nature.com/reprints. 
Correspondence and requests for materials should be addressed to Y.K. or H.M.Y.

\section*{Competing Financial Interests statement}
The authors declare no competing financial interests.

\begin{figure}[htbp]
  \begin{center}
    \includegraphics{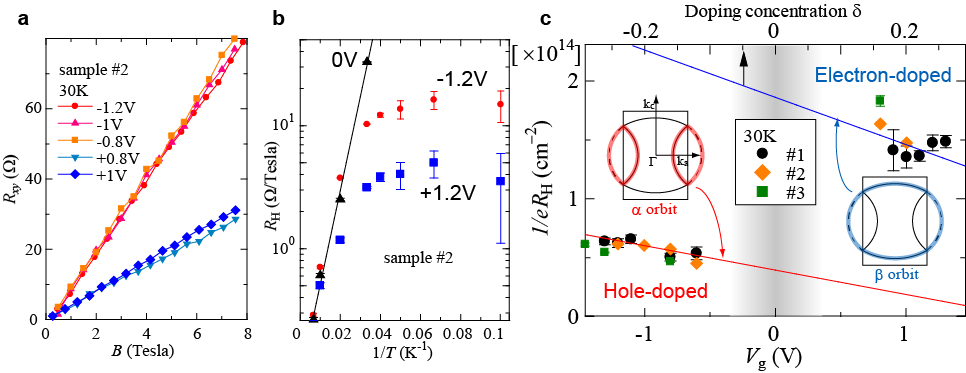}
  \caption{\textbf{Hall effect. a,} Hall resistance vs magnetic field at 30 K. \textbf{b,} Temperature dependence of $R_{\rm H}$ at $V_{\rm g}$ = $-$1.2, 0 and +1.2 V. The solid line represents the fit 8.78$\times 10^{-2} \times \exp{(E_{\rm a}/k_{\rm B}T)} \ \Omega$/Tesla with an activation energy $E_{\rm a}/k_{\rm B}$= 176 K to the data of the undoped bulk crystal. The error bars were calculated from the standard deviation of the Hall resistance vs magnetic field plots. Note that the data below 25 K contain further errors owing to the non-ohmic behaviour (Supplementary Note 6). \textbf{c,} Gate voltage dependence of $1/eR_{\rm H}$ at 30 K. Data from three different samples are shown. The solid lines denote the $1/eR_{\rm H}$ estimated from the volume bounded by the non-interacting FS (blue) and by the lenslike closed portion of the FS (red). Namely, the blue and red solid lines represent densities of the total carriers and the partial carriers enclosed by the lenslike closed portion, respectively. Here, the electron doping concentration $\delta$ is related to the non-interacting hole density $n$ (= 1$-\delta$) which is evaluated for the anisotropic Hubbard model with $U$ = 0. On the basis of the FET and EDLT measurements, the top horizontal axis is set in that $\delta$ = 0.2 corresponds to $V_{\rm g}$ = 1 V (see Supplementary Note 1).}
    \label{fig.3}
\end{center}
\end{figure}

\begin{figure}[htbp]
  \begin{center}
    \includegraphics{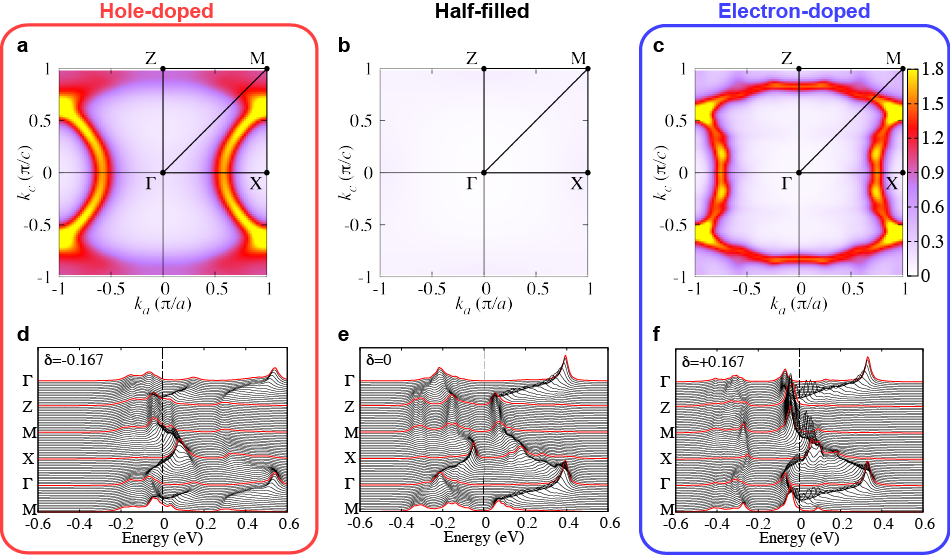}
  \caption{\textbf{Fermi surfaces and the single-particle spectral functions of the Hubbard model on an anisotropic triangular lattice at 30 K.} \textbf{a,b,c,} Fermi surfaces in the 1st BZ of $\kappa$-Cl for 17\% hole doping \textbf{(a)}, half filling \textbf{(b)} and 17\% electron doping \textbf{(c)}, determined by the largest spectral intensity at the Fermi energy. \textbf{d,e,f,} Single-particle spectral functions for 17\% hole doping \textbf{(d)}, half filling \textbf{(e)} and 17\% electron doping \textbf{(f)}. The Fermi energy is located at $\omega$ = 0 and the parameter set of this model is $t'/t$ = $-$0.8, $U/t$ = 7 and $t$ = 55 meV.}
    \label{fig.4}
\end{center}
\end{figure}

\begin{figure}[htbp]
  \begin{center}
    \includegraphics{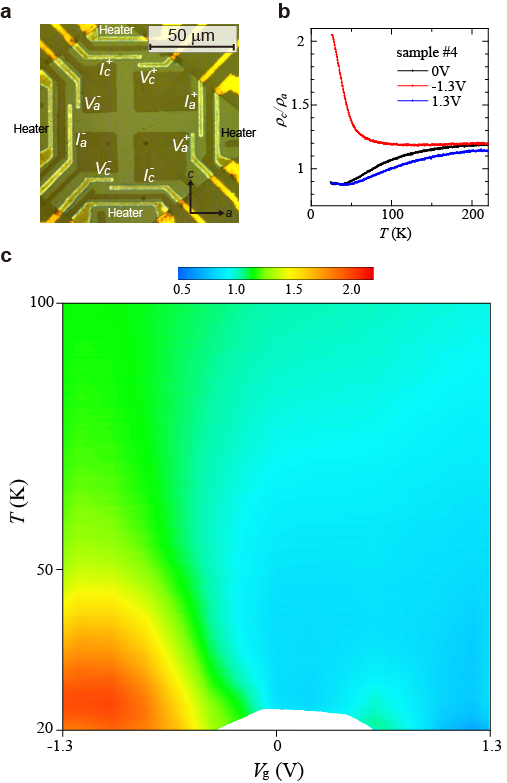}
  \caption{\textbf{Anisotropy of the resistivity a,} Optical image of the sample. The crystallographic axes were assigned by the heat measurement because the $a$-axis ($c$-axis) shows positive (negative) Seebeck coefficient at room temperature.  \textbf{b,} Temperature dependence of $\rho _{c}/\rho _{a}$ at $V_{\rm g}$ = $-$1.3, 0 and +1.3 V. Here, $\rho_ {c}$ ($\rho_ {a}$) is the resistivity along the $c$-axis ($a$-axis). \textbf{c,} Contour plot of $\rho _{c}/\rho _{a}$ as a function of $T$ and $V_{\rm g}$. The applied gate voltages were $-$1.3, $-$1.2, $-$1, $-$0.8, $-$0.6, $-$0.4, $-$0.2, 0, 0.2, 0.4, 0.6, 0.8, 1, 1.2 and 1.3 V. Data are missing at low temperature and low doping (white region) due to the high resistance.}
    \label{fig.5}
\end{center}
\end{figure}

\clearpage
\subsection*{Supplementary Figures}

\begin{figure}[htbp]
 \begin{center}
    \includegraphics{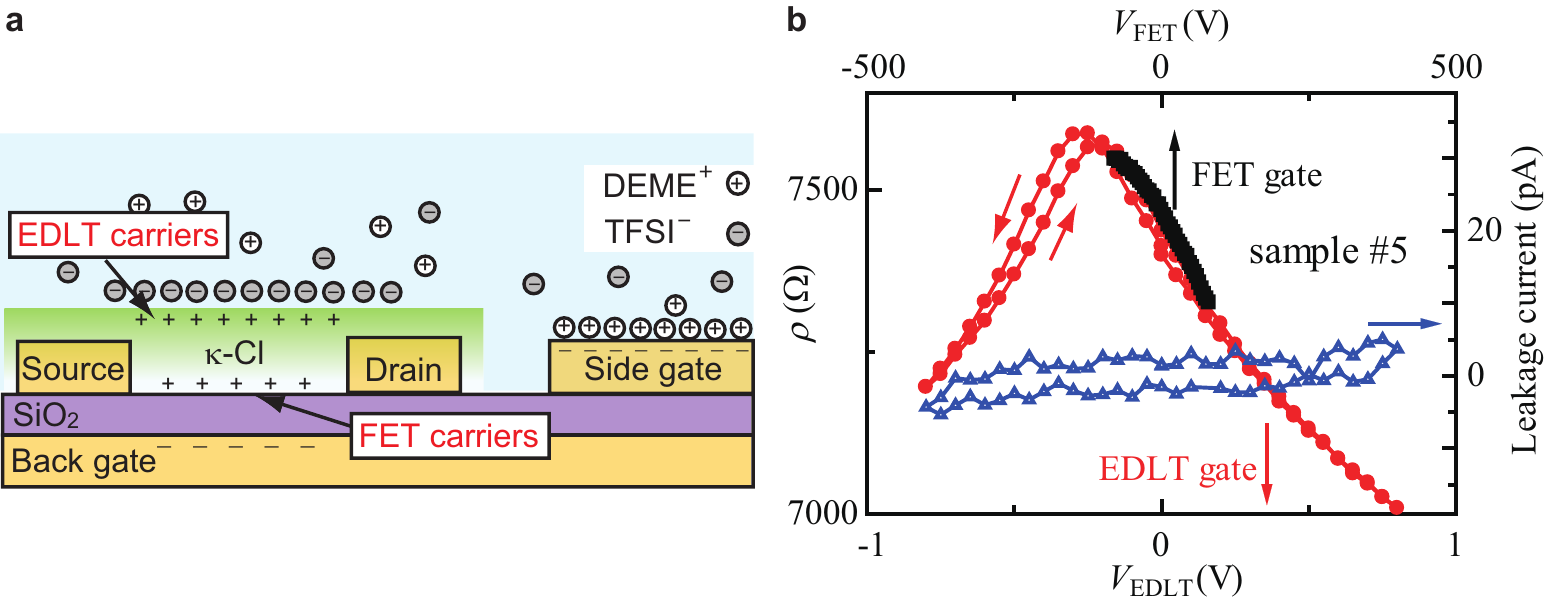}
    \caption{{\bf Comparison experiment between FET and EDLT devices using the same $\kappa$-Cl crystal. a,} Schematic side view of the FET and EDLT devices. {\bf b}, Transfer curves of sample \#5 at 220 K. The gate voltage dependence of the resistivity for the FET (black square) and EDLT operations (red circle) are shown. Blue triangle denotes the leakage current for the EDLT operation.}
\label{fig.s1}
\end{center}
\end{figure}

\begin{figure}[htbp]
  \begin{center}
    \includegraphics{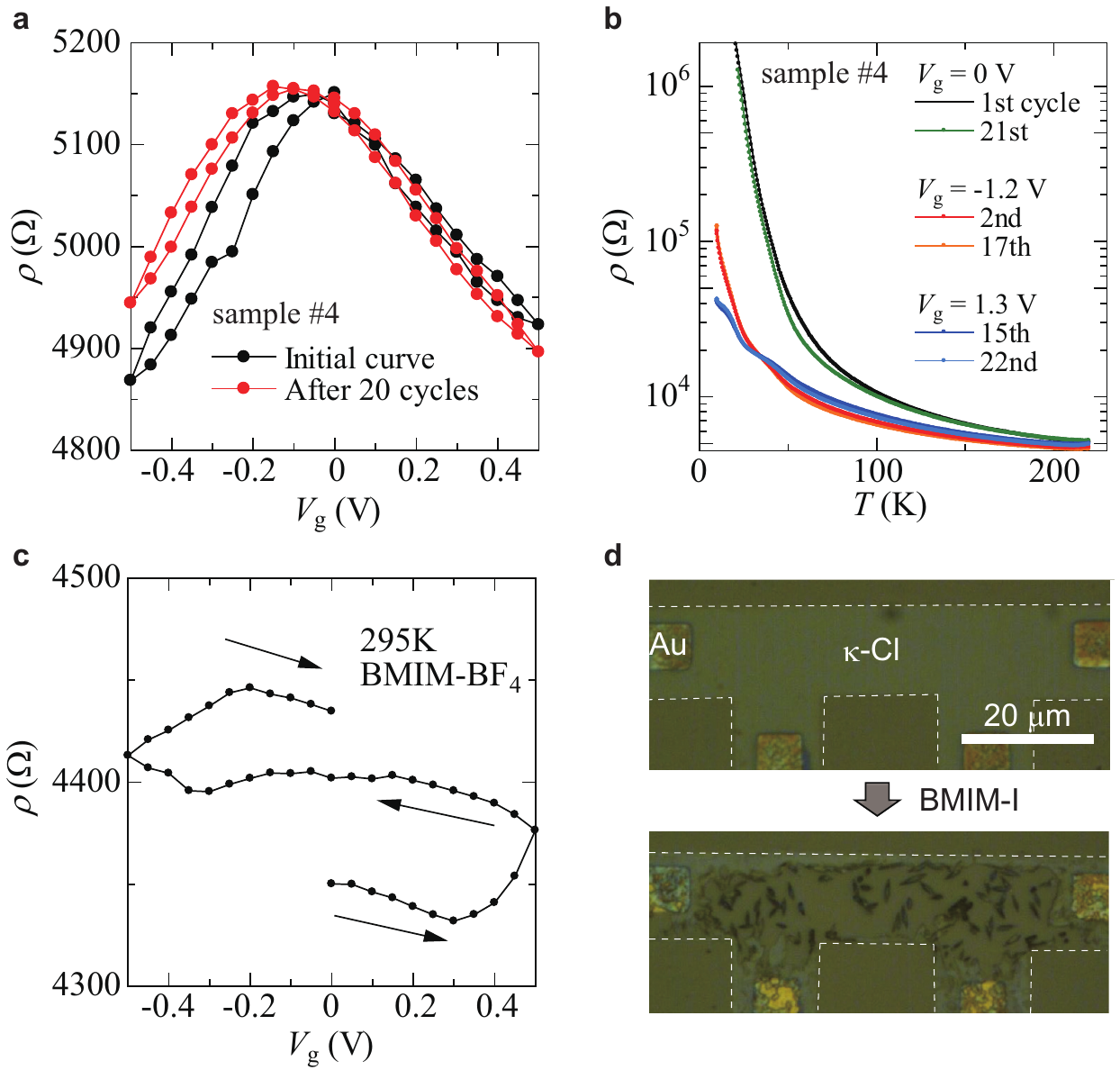}
  \caption{{\bf Repeatability of the measurements
  a,} Gate voltage dependence of the sheet resistivity at 220 K before and after 20 temperature cycles. {\bf b}, Temperature dependence of the sheet resistivity during the 1st, 21st, 2nd, 17th, 15th and 22nd temperature cycles. {\bf c}, Gate voltage dependence of the sheet resistivity at 295 K. {\bf d}, Optical image of a laser-shaped $\kappa$-Cl crystal before and after the drop of BMIM-I. The white broken lines denote the shape of the crystal. The sample was immediately rinsed with 2-propanol a few seconds after the drop.}
\label{fig.s2}
  \end{center}
\end{figure}

\begin{figure}[htbp]
  \begin{center}
    \includegraphics{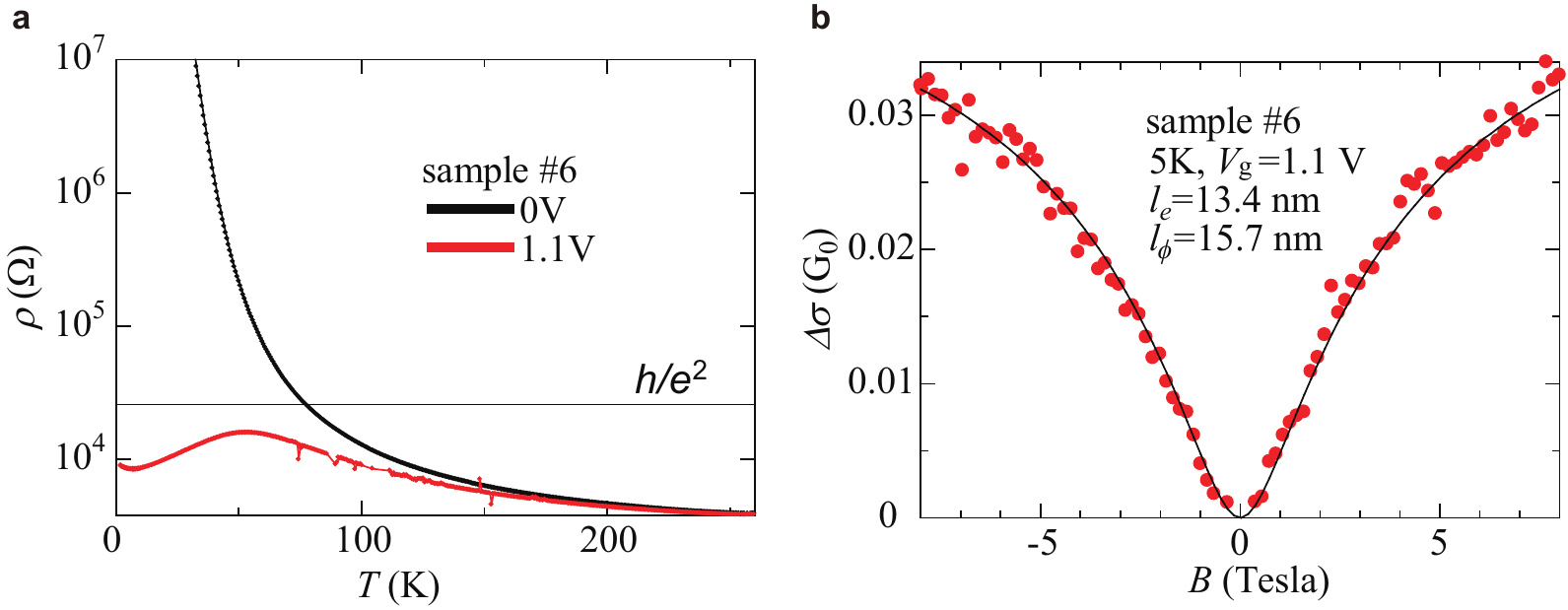}
  \caption{{\bf Negative magnetoresistance under electron doping. a,} Temperature dependences of the sheet resistivity in sample \#6 at $V_{\rm g}$ = 0 and +1.1 V. {\bf b}, Negative magnetoresistance at 5 K, $V_{\rm g}$ = +1.1 V where $G_{0} = 2e^{2}/h$ and least-squares fitting with the digamma functions (Supplementary Eq.(1)).}
\label{fig.s3}
  \end{center}
\end{figure}

\begin{figure}[htbp]
  \begin{center}
    \includegraphics{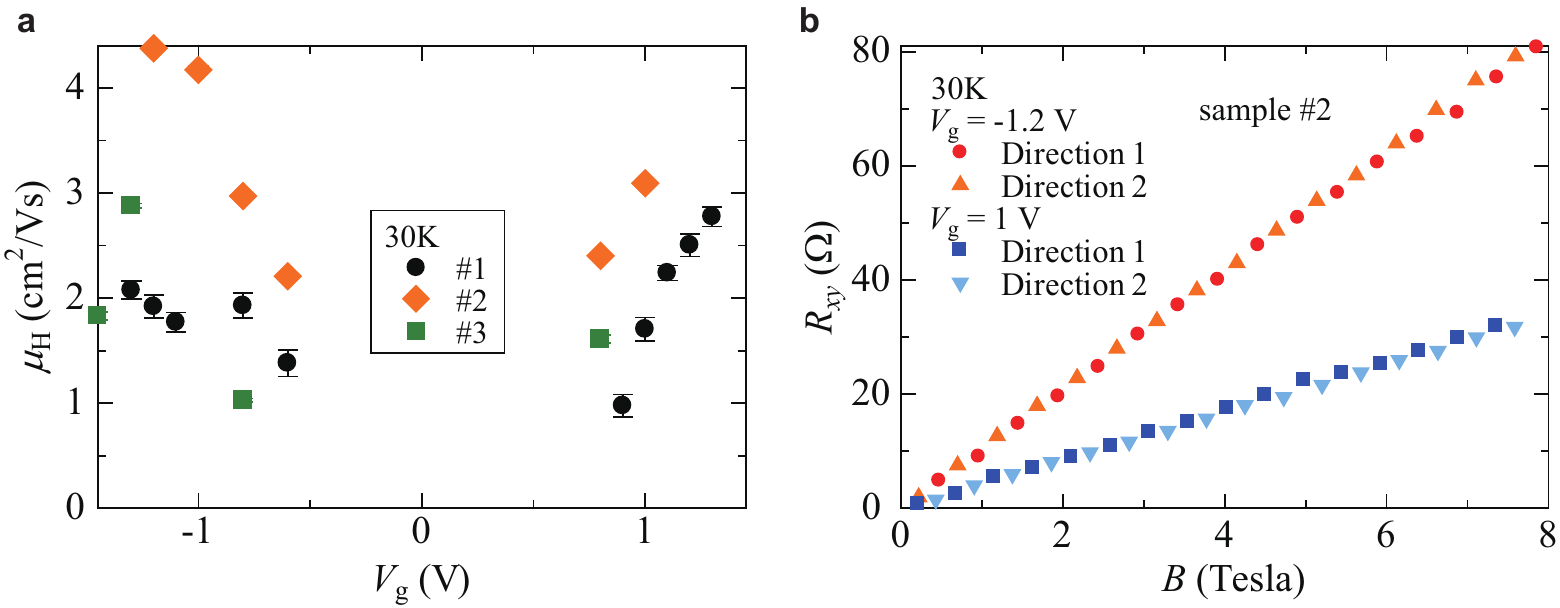}
  \caption{{\bf Hall mobility. a,} Gate voltage dependence of the Hall mobility at 30 K. The error bars were calculated from the standard deviation of the Hall resistance vs magnetic field plots. {\bf b}, Hall resistance vs magnetic field with two mutually-perpendicular current directions at 30 K. The directions 1 and 2 are parallel to the diagonals of the crystal, namely, $a$- or $c$-axis.}
\label{fig.s5}
  \end{center}
\end{figure}

\begin{figure}[htbp]
  \begin{center}
    \includegraphics{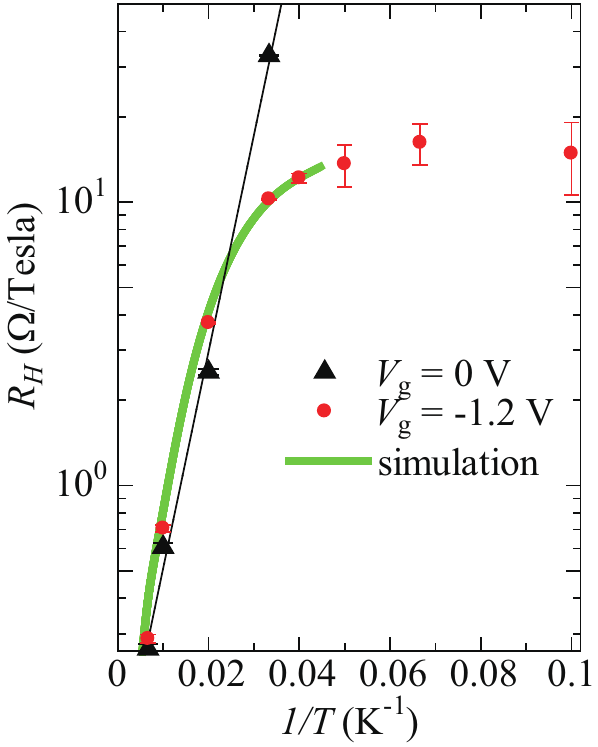}
  \caption{{\bf Temperature dependence of the measured and simulated Hall coefficient.} Black triangle and red circle denote the measured $R_{\rm H}$ at $V_{\rm g} = 0$ and $-$1.2 V, respectively. The measured $R_{\rm H}$ and the errors are taken from Fig.~\ref{fig.3}b in the main text. Green thick line is a simulation of $R_{\rm H}$ at $V_{\rm g} = -$1.2 V by Supplementary Eq.~(\ref{RHtotal}). It is reproduced by the constant $R_{\rm Hs}$ and thermally excited $R_{\rm Hb}$ (black thin line) down to approximately 22 K. At lower temperature, the simulation is not applicable because $\sigma _{\rm b}$ is not measurable.}
\label{fig.s6}
  \end{center}
\end{figure}

\begin{figure}[htbp]
  \begin{center}
    \includegraphics{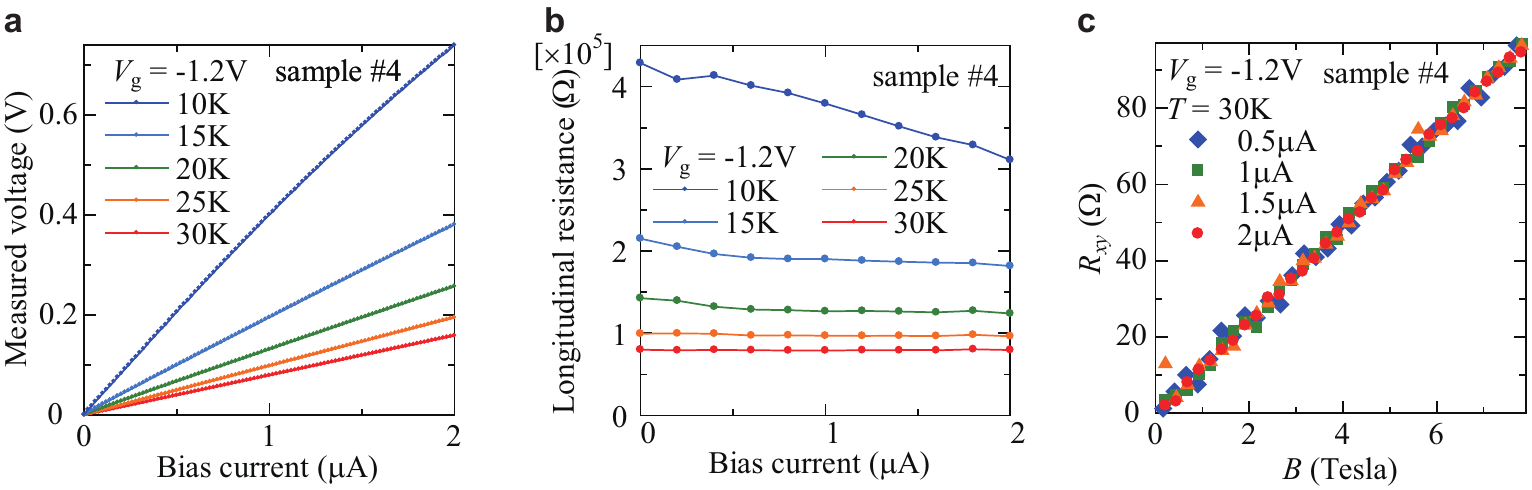}
  \caption{{\bf Current-voltage characteristics. a,b,} Current-voltage characteristics ({\bf a}) and current dependence of the longitudinal resistance ({\bf b}) at $V_{\rm g}$ = $-$1.2 V and $T = $ 10, 15, 20, 25 and 30 K. {\bf c}, Hall resistance vs magnetic field under various applied current at 30 K.}
\label{fig.s4}
  \end{center}
\end{figure}

\begin{figure*}png
  \begin{center}
    \includegraphics[width=16.0cm]{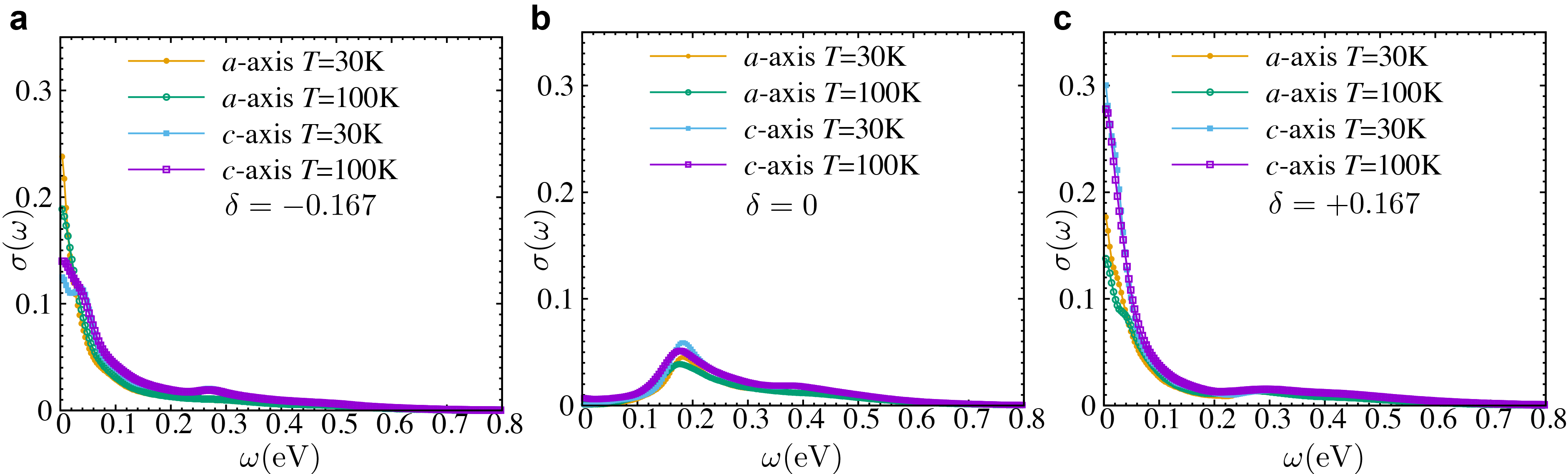}  
    \caption{            
      {\bf Optical conductivity.}
      The temperature dependence of the optical conductivity along the $a$- and $c$-axes for  
      {\bf a,} hole-doped ($\delta = -0.167$), 
      {\bf b,} half-filled ($\delta = 0$) and  
      {\bf c,} electron-doped ($\delta =  0.167$) cases.  
    }
    \label{fig.opt}
  \end{center}
\end{figure*}

\begin{figure*}
  \begin{center}
    \includegraphics[width=12.0cm]{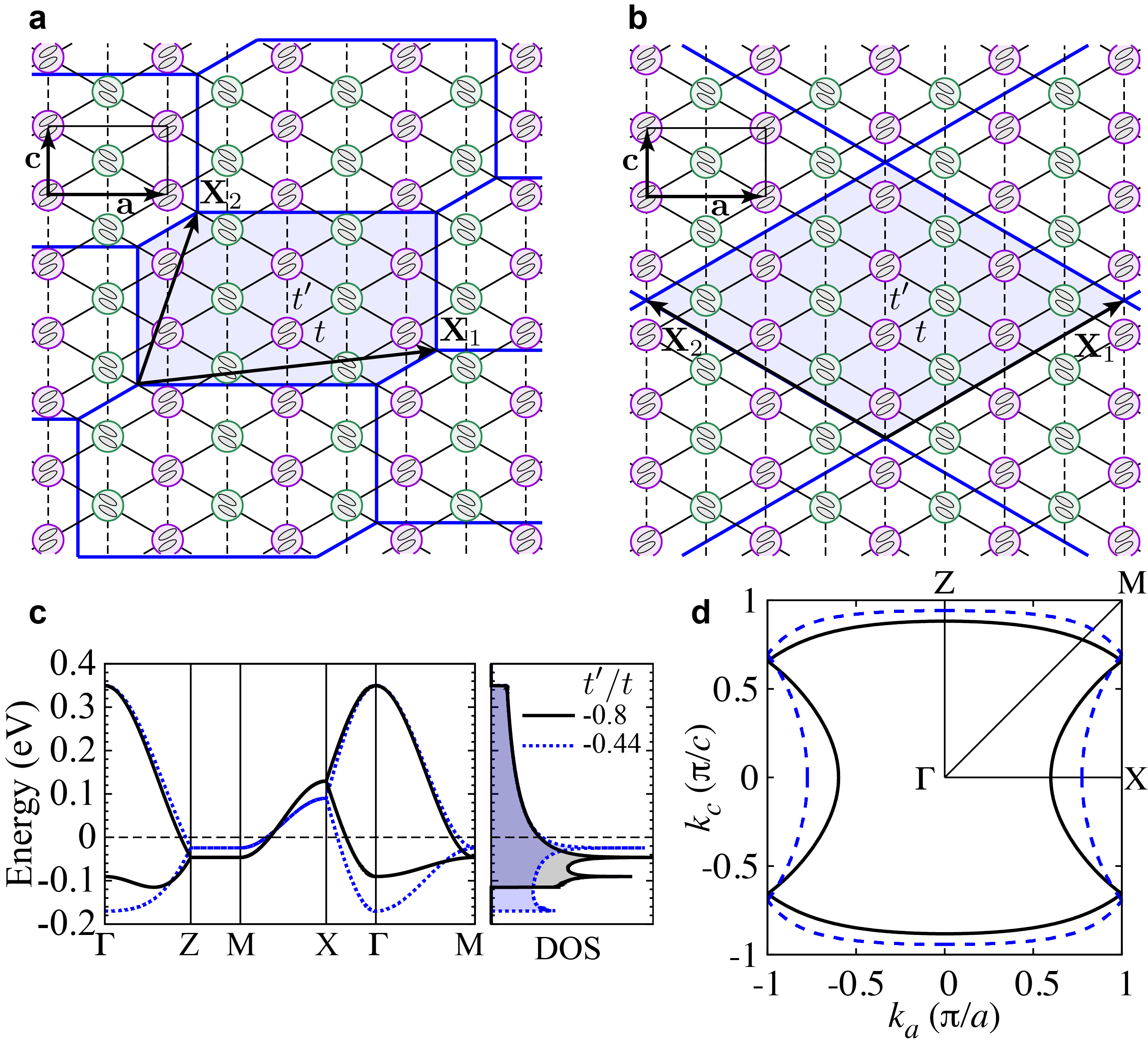}  
    \caption{
      {\bf The clusters used in the CPT calculations and electronic structures in the non-interacting limit. a,}
      A schematic of the BEDT-TTF layer ($a$-$c$ plane) of $\kappa$-type BEDT-TTF salts  
      described by the Hubbard model on the anisotropic triangular lattice 
      with transfer integrals $t$ (solid lines) and $t'$ (thin dashed lines).       
      The single site consists of a single BEDT-TTF dimer.
      Different orientation of dimers represents the crystallographically different dimers. 
      A unit cell and primitive translational vectors 
      $\mb{a} = (a,0)$ and $\mb{c} = (0,c)$ are indicated 
      as a solid rectangle and solid arrows, respectively. 
      The translational vectors $\mb{X}_1$ and $\mb{X}_2$ for the 12-site cluster (shaded area)
      are also denoted as thick solid arrows. 
      The blue solid lines represent the partitioning of the triangular lattice 
      into the 12-site clusters for the CPT calculations.
      {\bf b,}
      Same as {\bf a} but for the partitioning into the 16-site clusters.        
      {\bf c,}
      The non-interacting tight-binding band structure (left) 
      and the density of states (right) with 
      $t'/t = -0.8$ and $t = 55$ meV (solid lines). 
      For comparison, the results with $t'/t = -0.44$ and $t = 65$ meV are also shown by dashed lines.  
      The horizontal lines indicate the Fermi energy at half filling. 
      Here, $\Gamma = (0,0)$, Z$ = (0,\pi/c)$, M$ = (\pi/a,\pi/c)$, and X$ = (\pi/a,0)$.       
      {\bf d,}
      The Fermi surfaces for the non-interacting case with 
      $t'/t = -0.8$ (solid lines) and 
      $t'/t = -0.44$ (dashed lines) at half filling. 
    }
    \label{fig.lattice}
  \end{center}
\end{figure*}

\begin{figure*}
  \begin{center}  
    \includegraphics[width=16.0cm]{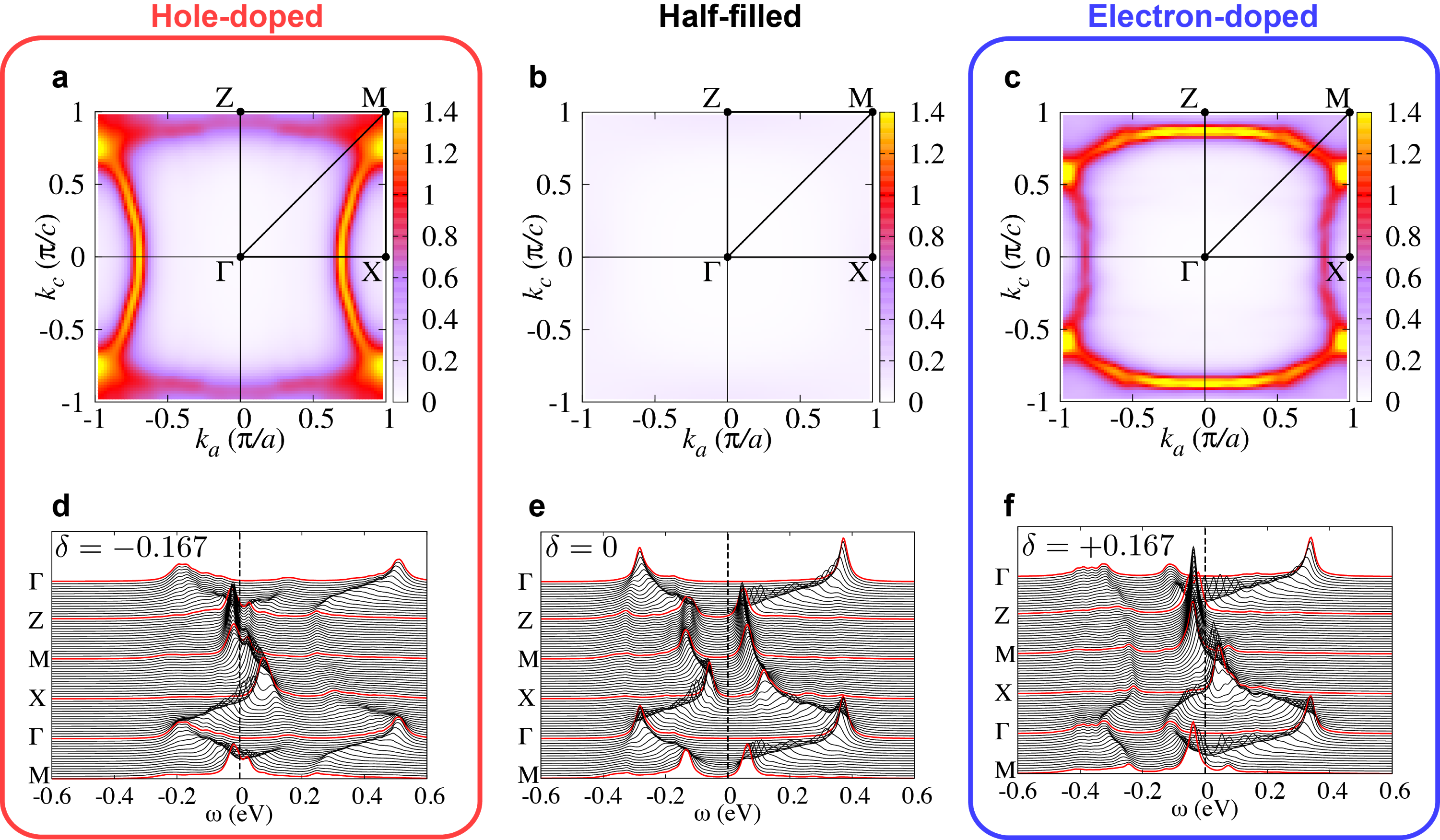}  
    \caption{
{\bf Fermi surfaces and the single-particle spectral functions of the Hubbard model on an anisotropic triangular lattice at 30 K.} \textbf{a,b,c,} Fermi surfaces in the 1st BZ of $\kappa$-Cl for 17\% hole doping \textbf{(a)}, half filling \textbf{(b)} and 17\% electron doping \textbf{(c)}, determined by the largest spectral intensity at the Fermi energy. \textbf{d,e,f,} Single-particle spectral functions for 17\% hole doping \textbf{(d)}, half filling \textbf{(e)} and 17\% electron doping \textbf{(f)}. The Fermi energy is located at $\omega$ = 0 and the parameter set of this model is $t'/t$ = $-$0.44, $U/t$ = 5.5 and $t$ = 65 meV.
}
    \label{fig.4another}
  \end{center}
\end{figure*}

\begin{figure*}
  \begin{center}
    \includegraphics[width=12.0cm]{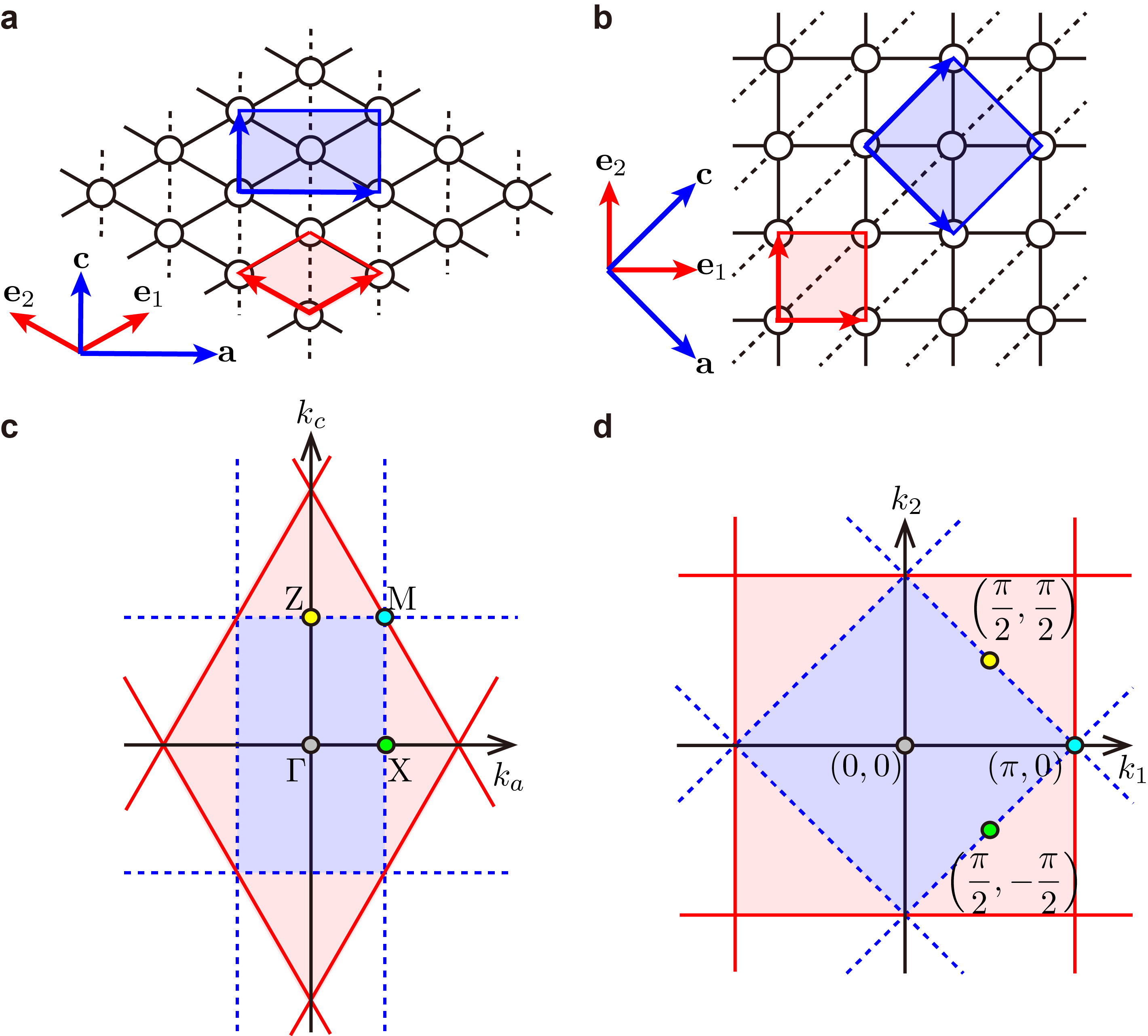}  
    \caption{             
      {\bf Unit cells and Brillouin zones. a,} 
      The anisotropic triangular lattice. 
      The translational vectors ${\mb{e}}_1$ and ${\mb{e}}_2$ (${\mb{a}}$ and ${\mb{c}}$)
      are represented by the red (blue) arrows. 
      The red (blue) shaded region represents the unit cell containing one site (two sites). 
      {\bf b,} 
      The lattice topologically equivalent to that in {\bf a}.
      Notice that ${\bf a}={\bf e}_{1}-{\bf e}_{2}$ and ${\bf c}={\bf e}_{1}+{\bf e}_{2}$ in both figures ${\bf a}$ and ${\bf b}$.  
      {\bf c,} The momentum space for the anisotropic triangular lattice in {\bf a}. 
      The BZ of the one- (two)-site unit cell is represented by 
      the red (blue) shaded region bounded by the red solid (blue dashed) lines. 
      {\bf d,} The same as {\bf c} but for the lattice in {\bf b}. 
    }
    \label{fig.BZ}
  \end{center}
\end{figure*}

\begin{figure*}
  \begin{center}  
    \includegraphics[width=16.0cm]{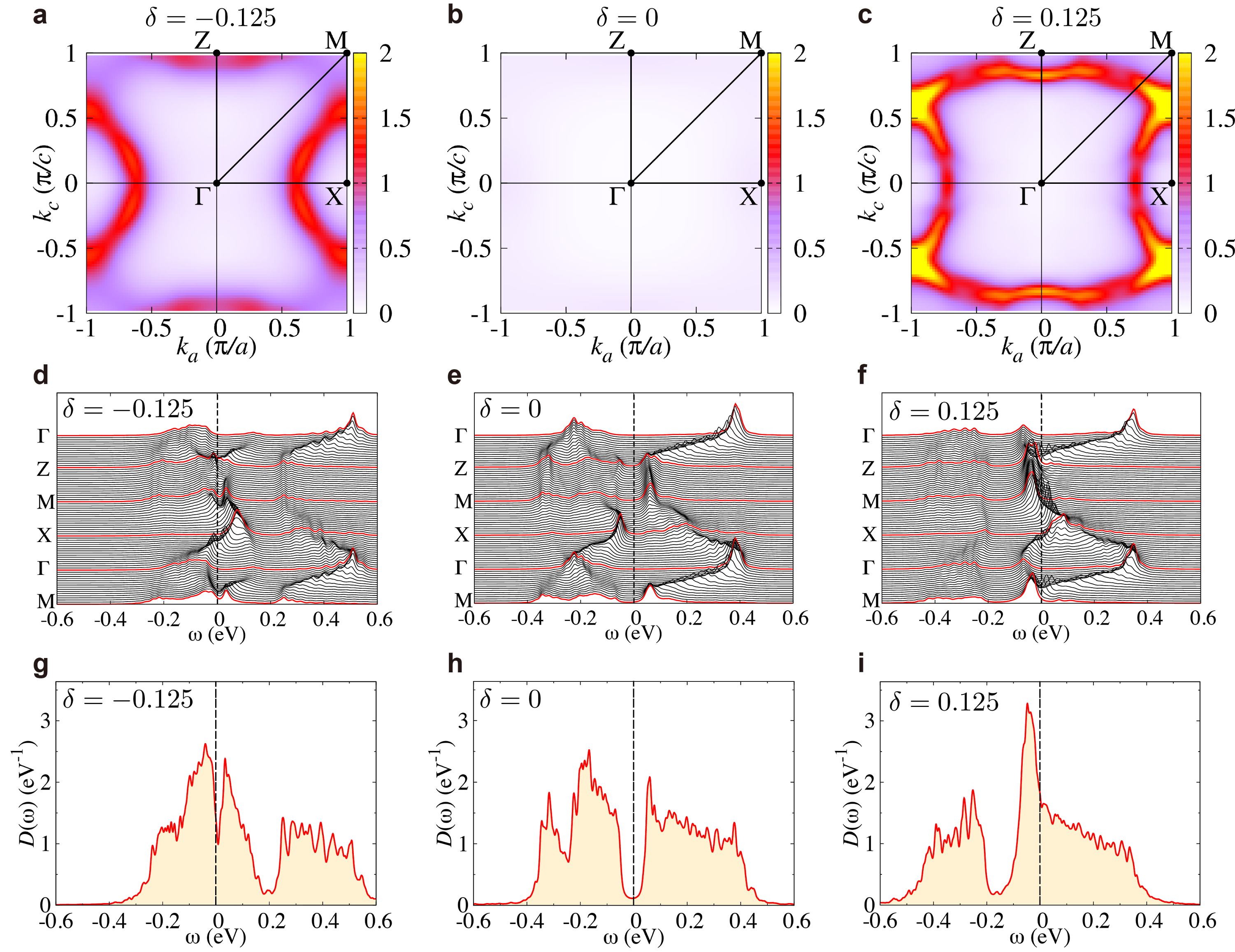}  
    \caption{
      {\bf Fermi surfaces, single-particle excitations, and density of states 
      of the single-band Hubbard model on an anisotropic  triangular lattice at 0 K with the 16-site cluster.}
      {\bf a, b, c,} The Fermi surfaces, 
      {\bf d, e, f,} the single-particle spectral functions and
      {\bf g, h, i,} the density of states 
      for hole-doped ({\bf a, d, g}), half-filled ({\bf b, e, h}) and electron-doped ({\bf c, f, i}) cases. 
      $\delta$ indicates the electron doping per dimer. 
      The Fermi energy is indicated as vertical dashed lines in d-i. 
    }
    \label{fig.Akw1}
  \end{center}
\end{figure*}

\begin{figure*}
  \begin{center}
    \includegraphics[width=16.0cm]{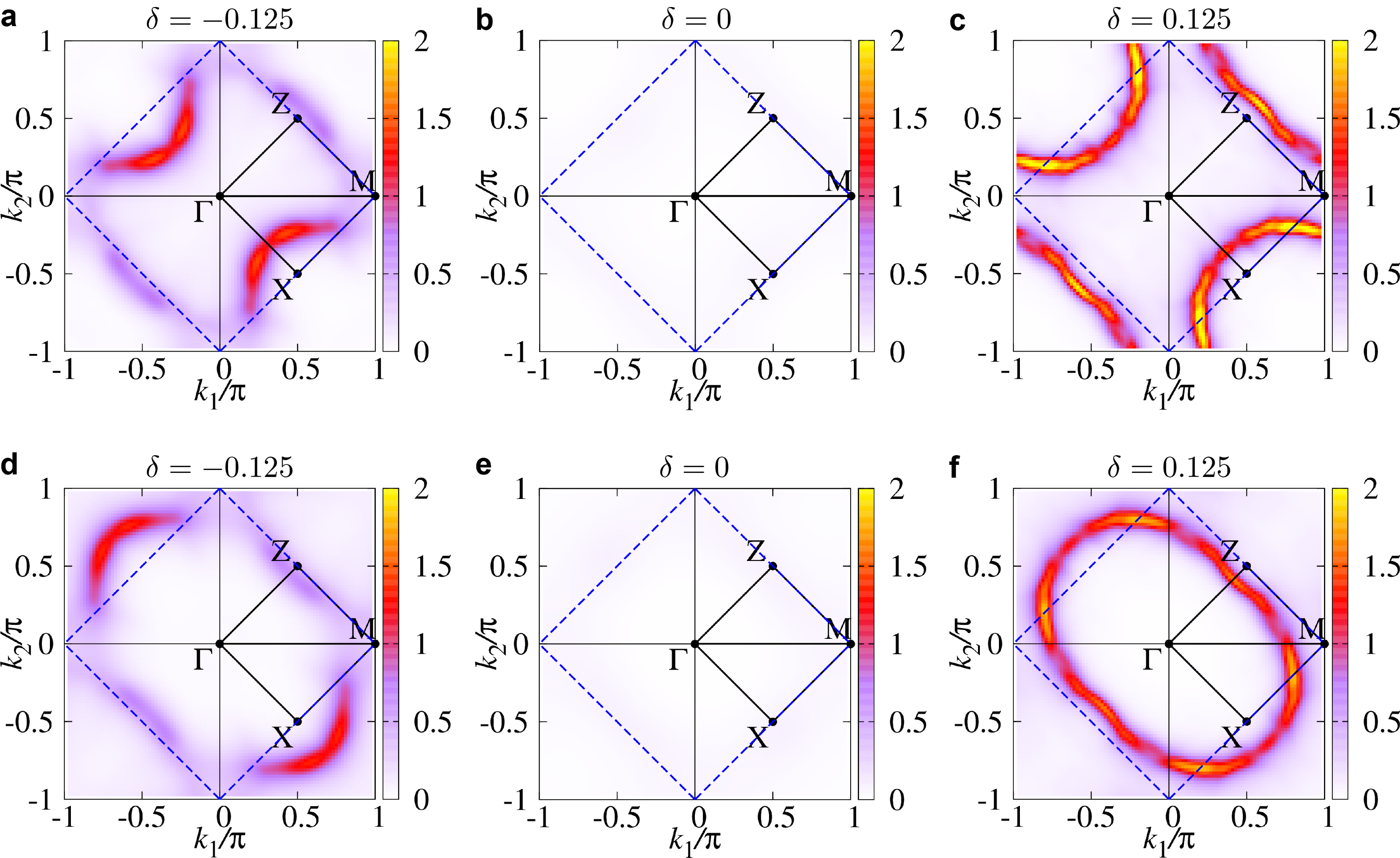}  
    \caption{
      {\bf  The Fermi surfaces in unfolded Brillouin zones at 0 K with the 16-site cluster.}             
      {\bf a, b, c,} The Fermi surface 
      at electron doping per dimer 
      $\delta = -0.125$ ({\bf a}), 
      $\delta = 0$ ({\bf b}), and 
      $\delta = 0.125$ ({\bf c})  
      shown in an ``unfolded" BZ.     
      {\bf d, e, f,} 
      Same as the top panels but for a different ``unfolded" BZ centered at 
      the $\Gamma$ point of the second BZ. 
      The wave vectors $k_1$ and $k_2$ are defined by Eq.(3) in the Methods section. 
      The blue dashed lines represent the boundaries of the BZ.   
    }
    \label{fig.Akw2}
  \end{center}
\end{figure*}

\clearpage
\subsection*{Supplementary Note 1: Comparison experiment between the FET and EDLT devices using the same $\kappa$-Cl crystal}
To check the carrier tunability of the electric double layer on $\kappa$-Cl, we compared the field effect of the EDLT (DEME-TFSI gate) and FET (SiO$_{2}$ gate) in the same $\kappa$-Cl crystal at 220 K (sample \#5, Fig.~\ref{fig.s1}).
Using the SiO$_{2}$ gate, we observed n-type FET behavior with a field-effect mobility of 2.6 cm$^{2}$/Vs.
On the other hand, a clear ambipolar field effect with a resistance peak at approximately $-$0.25 V was observed for the DEME-TFSI gate.
The hysteresis and the leakage current remained small without any signature of a chemical reaction between the electrolyte and $\kappa$-Cl.
As shown in Fig.~\ref{fig.s1}, the resistivity curves for the FET and EDLT almost coincide when we adjust the gate voltage axis.
Provided that the mobilities of the FET and EDLT are equivalent, the capacitance of the electric double layer on the electron-doping side is about 500 times larger than that of the FET. 
As a result, gate voltage of +1 V in the EDLT corresponded to approximately 20\% electron doping (3.6$\times$10$^{13}$ cm$^{-2}$).

As shown in Fig.~\ref{fig.3}c, the Hall coefficient at $V_{\rm g}$ = +1 V in sample \#2 (which shows the smallest error) gives a hole density of 1.47$\times$10$^{14}$ cm$^{-2}$ at 30 K. 
Using the half-filled hole density of $\kappa$-Cl, 1.86$\times$10$^{14}$ cm$^{-2}$, the density of injected electrons is estimated as 0.39$\times$10$^{14}$ cm$^{-2}$ ($=1.86\times$10$^{14}$ cm$^{-2}$$-$1.47$\times$10$^{14}$ cm$^{-2}$), which corresponds to 21\% electron doping. 
This value coincides the above estimate of 20\% within 1\% error. 
However, as stated in Supplementary Note 5, the experimental carrier density contains influences of thermally excited carriers in the bulk, and the absolute value of the above estimate of 20\% is not as accurate as it appears. 
Despite these complications, the relative trend of the carrier density should be reliable enough to support the discussions in the main text, because similar amount of errors in the same direction (or baseline shift) should be included for all the measured data.

\subsection*{Supplementary Note 2: Repeatability after low temperature measurements}
We checked the effect of multiple temperature cycles together with the gate voltage application, on the transfer curve at 220 K and the temperature dependence of the resistivity in sample \#4.
The applied gate voltage in each temperature cycle (between 220 and 10 K) was as follows: 0, $-$1.2, $-$1, $-$0.8, $-$0.6, $-$0.4, $-$0.2, 0, 0.2, 0.4, 0.6, 0.8, 1, 1.2, 1.3 1.4, $-$1.2, $-$1.3, 0, 1.3, 0, 1.3 and 1.35 V.
As shown in Fig.~\ref{fig.s2}a, the transfer curve at 220 K was reproducible after 20 temperature cycles despite the slight shift of approximately 0.1 V to the hole-doped side.
The temperature dependence of the resistivity (Fig.~\ref{fig.s2}b) was also reproducible after multiple temperature cycles with the gate voltage application of $\pm$1.3 V.
Therefore, it is unlikely that the sample is mechanically damaged due to thermal stress or degraded by destructive chemical reaction between the sample and gate electrolyte.

On the other hand, when we applied gate voltage at room temperature, the resistance irreversibly increased at $|V_{\rm g}|>0.3$ V (Fig.~\ref{fig.s2}c).
Moreover, when we dropped 1-butyl-3-methylimidazolium iodide (BMIM-I) on $\kappa$-Cl, the crystal was immediately dissolved without gate voltage as shown in Fig.~\ref{fig.s2}d.
Thus, although our device is weak against gate voltage application at room temperature and/or ionic liquids with large oxidizability, appropriate choice of ionic liquid and temperature enables repeatable measurements.

\subsection*{Supplementary Note 3: Negative magnetoresistance in a metallic sample under electron doping}
In a high-conductivity sample (sample \#6), we observed negative magnetoresistance at 5 K which is the hallmark of the weak localisation effect.
As shown in Fig.~\ref{fig.s3}, the negative magnetoresistance was fitted with the formula \cite{Nagaoka}
\begin{equation}
\Delta \sigma = -\frac{e^{2}}{\pi h}\left[ \psi \left( \frac{1}{2}+\frac{\hbar}{4eBl_{e}^{2}} \right)-\psi \left( \frac{1}{2}+\frac{\hbar}{4eBl_{\phi}^{2}} \right)-2{\rm ln}\frac{l_{\phi}}{l_{e}}\right]
\end{equation}
where $\psi $ denotes the digamma function, and $\Delta \sigma$, $B$, $l_{e}$ and $l_{\phi}$ are deviation of the sheet conductivity from that under zero magnetic field, the applied magnetic flux density, mean free path and dephasing length of the carriers, respectively.
$l_{e}$ and $l_{\phi}$ were estimated to be 13.4 nm and 15.7 nm from the least-squares fitting.
These values considerably exceed the distance between BEDT-TTF dimers ($\sim$1 nm) indicating coherent transport, in contrast to the Mott insulating state.

\subsection*{Supplementary Note 4: Gate voltage dependence of the Hall mobility}
The Hall mobility $\mu _{\rm H}$ is given by the product of the Hall coefficient and the conductivity.
Figure~\ref{fig.s5}a shows the gate voltage dependence of the Hall mobility in sample \#1$-$3 at 30 K.
The results have two implications: First, the Hall mobility is more sample-dependent than the Hall coefficient (Fig.~\ref{fig.3}c in the main text).
Namely, the mobility is highly sample dependent due to the surface conditions such as roughness, while the carrier density under gate voltage is more robust.
Second, the Hall mobilities are comparable between the hole- and electron-doped states, although the data contain ambiguities owing to the conductivity anisotropy (on the other hand, the Hall effect is isotropic as shown in Fig.~\ref{fig.s5}b).
This does not contradict to our calculations because the suppressed spectral function near the Z-M line under hole doping do not predominantly contribute to the Hall coefficient and conductivity.
Therefore, the difference of electron correlation effects between the electron- and hole-doped states cannot be examined via the Hall mobility in the present study.

\subsection*{Supplementary Note 5: Influence of thermally excited carriers on the Hall coefficient}
Since the samples consist of several tens of conducting BEDT-TTF layers, the $R_{\rm H}$ values more or less contain information of thermally excited carriers in the bulk. 
Here, we show that the presence of thermally excited carriers in the bulk explains the temperature dependence of $R_{\rm H}$ (Fig.~\ref{fig.3}b).
For sample $\#2$, we can roughly estimate the value at the surface (at $T$ = 30 K and $V_{\rm g} = -$1.2 V) on the assumption that $R_{\rm H}$ at $T$ = 30 K and $V_{\rm g}$ = 0 V purely originates from the bulk. 
When two types of hole carriers (for bulk and surface) coexist, the Hall coefficient $R_{\rm H}$ is expressed as
\begin{equation}
  R_{\rm H} 
  = \frac{\mu_{\rm b}^2 n_{\rm b} + \mu_{\rm s}^2 n_{\rm s}}{e(\mu_{\rm s} n_{\rm s} + \mu_{\rm s} n_{\rm s})^2} 
  = \frac{R_{\rm Hb} \sigma_{\rm b}^2 + R_{\rm Hs} \sigma_{\rm s}^2}{\sigma^2}
  \label{RHtotal}
\end{equation}
where $\mu_{\rm b/s}$, $n_{\rm b/s}$, $\sigma_{\rm b/s}=e \mu_{\rm b/s} n_{\rm b/s}$, and $R_{\rm Hb/s}=1/en_{\rm b/s}$ denote the mobility, carrier density, conductivity, and Hall coefficient for the bulk/surface, respectively, and $\sigma=\sigma_{\rm b} + \sigma_{\rm s}$.
The surface Hall coefficient $R_{\rm Hs}$ is therefore estimated from 
\begin{equation} 
  R_{\rm Hs} = \frac{R_{\rm H} \sigma^2 - R_{\rm Hb} \sigma_{\rm b}^2}{\sigma_{\rm s}^2}. 
  \label{RHsurface}
\end{equation}
By substituting measured values 
($R_{\rm H}$ $=10.25 \ \Omega/{\rm Tesla}$, 
$R_{\rm Hb}=32.92 \ \Omega/{\rm Tesla}$, 
$\sigma=43.40 \ \mu {\rm S}$, 
$\sigma_{\rm b}=8.56 \ \mu {\rm S}$ and 
$\sigma_{\rm s}=34.84 \ \mu {\rm S}$) 
at 30 K on the right-hand side of Supplementary Eq.~(\ref{RHsurface}), we obtain $R_{\rm Hs}=13.9 \ \Omega/{\rm Tesla}$ at $T=30$ K and $V_{\rm g}=-1.2$ V. 
Although this value contains ambiguity owing to the resistivity anisotropy, the value is close to that at 20 K (13.64 $\Omega/{\rm Tesla}$) and within the error bars of $R_{\rm H}$ at lower temperatures including $T=$ 20, 15 and 10 K.
The errors are calculated from the standard deviation of the $R_{xy}$ vs magnetic field plots and the ambiguity originating from the non-ohmic behaviour at low temperature (see Supplementary Note 6). 
Thus, the true value of $R_{\rm Hs}$ at 30 K is probably close to the measured $R_{\rm H}$ at 20 K, but the data below 30 K also contain non-ohmicity. 
We therefore used the data at 30 K in Fig.~\ref{fig.3}. 
Although we have such multiple origins of the errors on $R_{\rm H}$ value, the general trend of $R_{\rm H}$ vs $V_{\rm g}$ plot in the main text should not be as uncertain as the errors indicate, because those errors do not cause a random scattering but cause baseline shift which preserve the relative values between data points. 
Thus we believe that the $R_{\rm H}$ data are well representing the surface states and support our claims in the main text.

Having estimated the surface Hall coefficient $R_{\rm Hs}$ at 30 K, we then simulated the temperature dependence of the total Hall coefficient $R_{\rm H}$ at $V_{\rm g} = -1.2$ V with the two-carrier model in Supplementary Eq.~(\ref{RHtotal}), by assuming constant surface Hall coefficient $R_{\rm Hs} = 13.9 \ \Omega/{\rm Tesla}$ and thermally excited bulk Hall coefficient $R_{\rm Hb} = 8.78\times 10^{-2} \times \exp{(E_{\rm a}/k_{\rm B}T)}  \ \Omega/{\rm Tesla}$ with $E_{\rm a}/k_{\rm B} = 176$ K (black thin line in Fig.~\ref{fig.s6}).
The experimentally measured values are also used for the temperature dependence of $\sigma$, $\sigma_{\rm b}$, and $\sigma_{\rm s}$ in Supplementary Eq.~(\ref{RHtotal}). 
Notice that the simulation is not applicable for the temperatures lower than approximately 22 K because $\sigma_{\rm b}$ is not measurable. 
As shown in Fig.~\ref{fig.s6}, the measured values of $R_{\rm H}$ (red circle) are reproduced by this simple two-carrier model with constant surface Hall coefficient $R_{\rm Hs}$ and thermally excited bulk Hall coefficient $R_{\rm Hb}$ (green thick line). 
The increase in $R_{\rm H}$ upon cooling has been thus described by considering the influence of thermally excited carriers in the bulk.

\subsection*{Supplementary Note 6: Current-voltage characteristics in the insulating state under hole doping}
Since the hole-doped samples remained insulating, we measured the current-voltage characteristics at low temperature to test the ohmicity of the conduction.
Figures~\ref{fig.s4}a and b show the current-voltage characteristics and the current dependence of the longitudinal resistance at $V_{\rm g}$ = $-$1.2 V and $T = $ 10, 15, 20, 25 and 30 K. 
The current-voltage characteristics were non-ohmic below 20 K, whereas the longitudinal and Hall resistances did not depend on the applied current at 30 K (Fig.~\ref{fig.s4}b and c).

The error bars in Fig.~\ref{fig.3}b are calculated by taking into account the standard deviation $s$ of the $R_{xy}$ vs magnetic field plots and the ambiguity $\alpha$ originating from the non-ohmic behaviour. 
Since we applied 1 $\mu$A for the Hall measurements, $\alpha$ was defined as $\alpha = (R_{\rm 0A}-R_{\rm 1\mu A})/R_{\rm 1\mu A}$ at each temperature, where $R_{\rm 0A}$ and $R_{\rm 1\mu A}$ denote the resistances at 0 A and 1 $\mu$A in Fig.~\ref{fig.s5}b. 
The error bars in Fig.~\ref{fig.3}b were estimated from $(1+\alpha)(R_{\rm H}+s)-R_{\rm H}$ (here, $R_{\rm H}$ is the most probable value).

\subsection*{Supplementary Note 7: Optical conductivity} 
As described in the main text, we have observed experimentally that 
the electric resistivity exhibits anisotropy in particular 
for the hole-doped case at low temperature ($T \sim 30$ K), and 
the anisotropy is gradually weakened with increasing the temperature. 
Here, we calculate the optical conductivity along $a$- and $c$-axes and show that they indeed 
exhibits the anisotropy and the temperature dependence consistent with 
the experiments. 

The optical conductivity along $a$ ($c$) direction is defined as 
\begin{equation}\label{sigma}
  \s_{a/c}(\w) = 
  \frac{\pi}{V_{\rm BZ}} \int _{\rm BZ} \dd \mb{k} \int \dd \w' \frac{f(\w'-\w) - f(\w')}{\w} A(\mb{k},\w'-\w) A(\mb{k},\w') v_{\mb{k}a/c}^2,   
\end{equation}
where $f(\w) = (\e^{\w/T }+1)^{-1}$ is the Fermi distribution function, 
\begin{equation}
  v_{\mb{k},a} = t \sin{k_1} -t \sin{k_2} 
\end{equation}  
is the velocity of the non-interacting electrons along $a$-axis, and 
\begin{equation}
  v_{\mb{k},c} = t \sin{k_1} +t \sin{k_2} + 2t' \sin(k_1 + k_2)   
\end{equation}  
is that along $c$-axis. Notice that the optical conductivity in Supplementary Eq.~(\ref{sigma}) 
does not include the vertex corrections. 
Figures~\ref{fig.opt}{\bf a}, {\bf b}, and {\bf c} show the temperature dependence of 
the optical conductivity along the $a$- and $c$-axes for hole-doped, half-filled, and electron-doped cases, respectively. 
Here we use the parameter set of $t'/t = -0.44$, $U/t = 5.5$, and $t = 65$ meV.  

At half filling, a broad peak appears at $\w \sim 0.18$ eV. 
This peak corresponds to the excitation from 
the lower Hubbard band to the upper Hubbard band.
Although the peak is slightly broadened when the temperature is increased, 
its position and overall structure are almost independent of the temperature 
since the energy scale of the electron interaction $U \simeq 0.36$ eV $\simeq 4200$ K 
is much larger than that of the temperature considered.  
In the doped cases, the peak at $\w \sim 0.18$ eV found at half filling disappears 
while a shoulder remains at $\w \sim 0.25$ eV. 
This spectral-weight transfer is the characteristic feature of the doped Mott insulators~\cite{Eskes1994} 
also observed in the cuprate high-temperature superconductor~\cite{Uchida1991}.

As shown in Fig.~\ref{fig.opt}{\bf b}, the anisotropy is weak at half filling.
This is because the high-energy excitations across the Mott gap 
originates in the local electronic states, 
and therefore it is insensitive to the direction.  
When the carriers are doped, the zero-energy excitations 
become possible and the anisotropy is clearly observed 
in the low-energy region ($\w \lesssim 0.05$ eV).  
For the hole doped case (Fig.~\ref{fig.opt}{\bf a}), 
$\s_a(\w)$ is twice as large as $\s_c(\w)$ at $T = 30$ K in the low-energy limit, 
but the anisotropy becomes weaker when the temperature is increased to $T = 100$ K.
For the electron doped case at $T = 30$ K, 
the anisotropy is smaller than that in the hole-doped case. 
However, as shown in Fig.~\ref{fig.opt}{\bf c}, the anisotropy is reversed under the electron doping. 
This is in good accordance with the resistivity measurement under the electron doping at low temperatures. 
We have thus shown that the low-energy excitations in the optical conductivity 
exhibit significant spatial anisotropy and temperature dependence, 
which is in good qualitative agreement with the resistivity experiment.

\subsection*{Supplementary Note 8: Additional results at zero temperature with the 16-site cluster}
Figure~\ref{fig.Akw1} summarizes the results of the 16-site cluster 
with $t'/t = -0.8$, $U/t = 7$, and $t = 55$ meV calculated at zero temperature. 
The electron-hole asymmetry of the Fermi surface can be seen clearly in Fig.~\ref{fig.Akw1}a-c. 
The single-particle spectral functions in Fig.~\ref{fig.Akw1}d-f 
exhibit qualitatively the same results 
with those of the 12-site cluster shown in the main text. 
Fig.~\ref{fig.Akw1}g-i show the density of states 
\begin{equation}
  D(\w) =  \frac{1}{V_{\rm BZ}} \int_{\rm {BZ}}\dd \mb{k} A(\mb{k},\w), 
\end{equation}
where $V_{\rm BZ} = (2\pi)^2/(ac) $ is the volume of the BZ and 
the integration over $\mb{k}$ is taken in the BZ.
The substantial suppression of the density of states at the Fermi energy, 
i.e., the pseudogap, is observed in the hole-doped case (see Fig.~\ref{fig.Akw1}g), 
consistent with the observation of the Fermi arc. 
On the other hand, such suppression of the density of states at 
the Fermi energy is hardly seen in the electron doped case (see Fig.~\ref{fig.Akw1}i). 
These results demonstrate that 
the electron-hole asymmetry is not affected by the clusters chosen for the CPT calculations.

Finally, we show the Fermi surfaces in ``unfolded" BZ in Fig.~\ref{fig.Akw2}. 
These plots are easily compared with other theoretical calculations 
often performed on the triangular lattice with the single-site unit cell~\cite{Kang2011}.  
Fig.~\ref{fig.Akw2}a-c 
show the Fermi surface calculated with the 16-site cluster at zero temperature. 
Here the unfolded BZ is centred at $\Gamma$ point of the first BZ of the triangular lattice 
with the two-site unit cell. 
Fig.~\ref{fig.Akw2}d-f are the same as panels a-c, 
but the unfolded BZ is centered at  $\Gamma$ point of one of the second BZs. 
In both plots, one can again observe clearly the electron-hole asymmetric reconstruction of the 
Fermi surface.

\end{document}